\documentclass[alpha-refs]{wiley-article}

\pdfoutput=1
\usepackage{graphicx}
\usepackage[space]{grffile}
\usepackage{latexsym}
\usepackage{textcomp}
\usepackage{longtable}
\usepackage{tabulary}
\usepackage{booktabs,array,multirow}
\usepackage{amsfonts,amsmath,amssymb}
\usepackage{natbib}
\usepackage{url}
\usepackage{hyperref}
\hypersetup{colorlinks=false,pdfborder={0 0 0}}
\usepackage{etoolbox}

\newif\iflatexml\latexmlfalse

\AtBeginDocument{\DeclareGraphicsExtensions{.pdf,.PDF,.eps,.EPS,.png,.PNG,.tif,.TIF,.jpg,.JPG,.jpeg,.JPEG}}

\usepackage[utf8]{inputenc}
\usepackage[english]{babel}

\usepackage{siunitx}

\iflatexml

\else

\paperfield{Field of the paper}
\corraddress{M. A. C. Teixeira PhD, Department of Meteorology, University of Reading, Reading, Berkshire, RG6 6ET, UK}
\corremail{m.a.teixeira@reading.ac.uk}
\fi

\papertype{Original Article}

\title{The drag exerted by weakly dissipative trapped lee waves on the atmosphere: application to Scorer's two-layer model}

\author[1]{Miguel A. C. Teixeira}
\author[2,3]{Jos\'e L. Arga\'\i n}

\affil[1]{Department of Meteorology, University of Reading, Reading, UK}
\affil[2]{Department of Physics, University of Algarve, Faro, Portugal}
\affil[3]{CIMA, University of Algarve, Faro, Portugal}

\runningauthor{M. A. C. Teixeira and J. L. Arga\'\i n}

\begin{document}

\maketitle
\selectlanguage{english}
\begin{abstract}
While it is known that trapped lee waves propagating at low levels in a stratified atmosphere exert a drag on the mountains that generate them, the distribution of the corresponding reaction force exerted on the atmospheric mean circulation, defined by the wave momentum flux profiles, has not been established, because for inviscid trapped lee waves these profiles oscillate indefinitely downstream.
A framework is developed here for the unambiguous calculation of momentum flux profiles produced by trapped lee waves, which circumvents the difficulties plaguing the inviscid trapped lee wave theory. Using linear theory, and taking Scorer's two-layer atmosphere as an example, the waves are assumed to be subject to a small dissipation, expressed as a Rayleigh damping. The resulting wave pattern decays downstream, so the momentum flux profile integrated over the area occupied by the waves converges to a well-defined form. Remarkably, for weak dissipation, this form is independent of the value of Rayleigh damping coefficient, and the inviscid drag, determined in previous studies, is recovered as the momentum flux at the surface. The divergence of this momentum flux profile accounts for the areally-integrated drag exerted by the waves on the atmosphere.
The application of this framework to this and other types of trapped lee waves potentially enables the development of physically-based parametrizations of the effects of trapped lee waves on the atmosphere.

\end{abstract}

\section{Introduction}
\label{intro}

Orographic internal gravity waves in a stratified atmosphere (also known as mountain waves) exert a pressure drag on the mountains that generate them. By Newton's third law, a reaction force of equal magnitude and opposite direction must be exerted by the mountains on the atmosphere \citep{Nappo_2012}. Since air is a fluid, this reaction force may be distributed spatially, in some cases over large distances, either vertically or horizontally. The mountains that act as a source of these waves have typical widths of order 10 km or smaller, and usually are not resolved explicitly in weather prediction or climate models, so the waves must be parametrized \citep{Stensrud_2009}. The total value of the orographic gravity wave drag and its spatial distribution need to be specified in such parametrizations (e.g. \citealp{Lott_Miller_1997}).

A theory for the generation and dissipation of hydrostatic gravity waves, which propagate vertically in the atmosphere, has been extensively developed over the last decades (\citealp{Phillips_1984}; \citealp{McFarlane_1987}; \citealp{Shutts_1995}; \citealp{Shutts_Gadian_1999}), and serves as the physical basis for existing orographic gravity wave drag parametrizations. In this theory, where the linear approximation is typically made, the total drag can often be calculated analytically (\citealp{Smith_1979a}; \citealp{Phillips_1984}; \citealp{Teixeira_etal_2004}; \citealp{Teixeira_Miranda_2006}), and even the distribution of the force exerted on the atmosphere, which is specified through the divergence of the vertical flux of horizontal wave momentum, can sometimes be expressed analytically (\citealp{Shutts_1995}; \citealp{Shutts_Gadian_1999}; \citealp{Teixeira_Miranda_2009}; \citealp{Teixeira_Yu_2014}). This tractability results from the simplifications inherent to the linear and hydrostatic approximations, whereby the waves not only are always vertically propagating, but are also absorbed by critical levels in an inviscid context via a mechanism that mimics their more realistic (finite-amplitude) attenuation by wave breaking (\citealp{Booker_Bretherton_1967}; \citealp{Grubisic_Smolarkiewicz_1997}; \citealp{Shutts_1995}; \citealp{Teixeira_etal_2008a}). For stationary waves, critical levels can be defined as levels in the atmosphere where the mean wind velocity is perpendicular to the horizontal wavenumber vector of the wave, or is simply zero. Since the mean wind vector is likely to turn substantially with height, or vanish, over the depth of the atmosphere, critical levels are an effective mechanism for momentum transfer from the waves to the mean flow. Other mechanisms that lead to an increase in the amplitude of the waves as their energy propagates upward, and therefore to their breaking and dissipation, with momentum transfer to the mean flow, are the decrease of density and variation of static stability with height (\citealp{Smith_1979a}; \citealp{McFarlane_1987}). In hydrostatic waves, which propagate essentially vertically, the occurrence of any of these factors in an atmospheric column over the source orography will ensure that the wave momentum flux is totally deposited into the mean flow as the reaction force acting on the atmosphere.

However, the situation is less clear for trapped lee waves, and non-hydrostatic waves in general, whose properties have received much less attention \citep{Xu_etal_2021}. Trapped lee waves are intrinsically non-hydrostatic mountain waves that propagate horizontally in the atmosphere, as a result of vertical reflection and trapping (leading to ducting) within a layer, typically adjacent to the ground (\citealp{Scorer_1949}; \citealp{Vosper_etal_2006}). Analytical expressions for the total drag produced by these waves have been derived and tested, both for generic cases (\citealp{Bretherton_1969}; \citealp{Smith_1976}; \citealp{Gregory_etal_1998}) and for waves propagating in simple two-layer atmospheres (\citealp{Teixeira_etal_2013a},\citeyear{Teixeira_etal_2013b}; \citealp{Teixeira_Miranda_2017}). These waves are expected to be dissipated primarily via friction within the boundary layer, as their energy repeatedly propagates towards the ground and is reflected by it (\citealp{Jiang_etal_2006}; \citealp{Lott_2007}), but there is no clear idea of how the divergence of the wave momentum flux may exert drag on the mean flow in that case. The reason is that, unlike in inviscid linear hydrostatic theory, where critical levels or the decay of density with height provide natural ways of producing a momentum flux divergence, there is no such mechanism for waves that propagate horizontally. Additionally, the inviscid solution from linear theory for trapped lee waves produces momentum fluxes that are both horizontal and vertical \citep{Broad_2002} and that are ill-defined, oscillating with the wave phase of the (horizontally infinite) wave train. There have been attempts to analyse the impact of trapped lee waves on the atmosphere with recourse to the concept of wave pseudo-momentum \citep{Shepherd_1990}, but progress has been limited by the fact that a non-dissipative framework was adopted (\citealp{Durran_1995}; \citealp{Lott_1998}).

As will be seen in the present study, in order to obtain a well-posed mathematical problem for the trapped lee waves (even restricted to linear theory) that allows a derivation of the effect of the waves on the mean flow through the momentum flux divergence terms in the equations of motion, it is necessary to introduce at least weak dissipation.
The corresponding treatment provides an example of a situation in a fluid flow in which the limit of the solution when friction approaches zero is different from the solution when friction is assumed from the outset to be exactly zero, and physically meaningful results are only obtained in the former case. This parallels the mechanisms in boundary layer theory that resolve D'Alembert's paradox, and in other fluid dynamics problems involving the effects of weak friction (e.g. \citealp{Teixeira_etal_2012}).

In this study, the simplest representation of friction as a Rayleigh damping will be adopted, and results will be illustrated for the case of the two-layer atmosphere of Scorer (\citealp{Scorer_1949}; \citealp{Teixeira_etal_2013a}), but the results are found to be independent of the value of the Rayleigh damping coefficient, as long as this is small, and the concept underlying the calculations appears to be generalizable to other model atmospheres. The independence of the results from the details of the Rayleigh damping suggest, in particular, that they may be independent of the type of dissipation adopted (as long as this is weak), and probably constitute the true quasi-inviscid solution to the momentum flux profiles that may serve as a leading-order orographic forcing in gravity wave drag parametrizations.

This paper is organized as follows: Section 2 describes theoretical developments, including an extension of inviscid results, results with weak friction, and their application to Scorer's atmosphere. Section 3 presents the nonlinear numerical model and the linear model with friction against which the theory is compared. Section 4 presents some preliminary comparisons, both purely inviscid and with vanishing friction, used to validate the theoretical results. Finally, Section 5 summarizes the main conclusions of this study.

\section{Theory}
\label{method}

In view of the difficulties pointed out above, the existing theory for the momentum fluxes associated with trapped lee waves
can be considered unsatisfactory and
incomplete.
Since basic aspects still need attention, the present treatment will be limited to conditions under which two-dimensional (2D) linear theory is valid, and a brief review of previous results is included in the theoretical development.

For vertically propagating waves, the effect of the waves on the mean flow (which in a parametrization corresponds to the resolved atmospheric circulation) is given by (cf. \citealp{Stensrud_2009}; \citealp{Nappo_2012})
\begin{equation}
\rho \frac{\partial \langle U \rangle}{\partial t} = -\frac{\partial}{\partial z} \left( \rho \langle u w \rangle \right) + {\rm other \,\, terms},
\label{tendency}
\end{equation}
where $U$ is the mean wind velocity
(in the $x$ direction), $u$ and $w$ are the horizontal and vertical velocity perturbations associated with the waves and $\rho$ is the density. The angle brackets denote the average over a certain area or (in 2D) spatial distance along $x$, say:
\begin{equation}
\langle uw \rangle = \frac{\int_{-\frac{\Delta x}{2}}^{+\frac{\Delta x}{2}} uw \, dx}{\Delta x}.
\label{average}
\end{equation}
In Equation (\ref{tendency}), the nonlinear term explicitly presented on the right-hand side inside the $z$ derivative is the wave momentum flux, which causes a deceleration (or acceleration) of the mean flow. In Equation (\ref{average}), $\Delta x$ may represent, for example, the grid spacing along $x$ in the model where the drag parametrization is implemented. Implicit in the terms omitted in Equation (\ref{tendency}) is the idea that the contribution to the drag from the divergence of the horizontal momentum fluxes is irrelevant, as those fluxes become zero at the edges of the integration domain used in Equation (\ref{average}). This is consistent with vertically propagating (hydrostatic) waves generated by an isolated mountain, which justify a so-called `single-column' approach to drag parametrization.  

In the theory of internal gravity waves generated by isolated mountains that serves as a basis for most drag parametrizations, what is called the wave momentum flux is often denoted by
\begin{equation}
M=\rho \int_{-\infty}^{+\infty} u w \, dx,
\label{momflux}
\end{equation}
or the corresponding 2D version (\citealp{Bretherton_1969}; \citealp{Shutts_1995}; \citealp{Teixeira_Miranda_2009}; \citealp{Teixeira_Yu_2014}). The definition of Equation (\ref{momflux}) will be adopted here.
Despite the fact that the integration limits in Equation (\ref{momflux}) are different from those in Equation (\ref{average}), if the waves are hydrostatic and generated by an isolated mountain, the integral should take the same value.
From inviscid linear wave theory, it can be shown (\citealp{Smith_1979a}; \citealp{Nappo_2012}; \citealp{Teixeira_Miranda_2009}) that
\begin{equation}
M(z=0)=\rho \int_{-\infty}^{+\infty} uw(z=0) \, dx = -\int_{-\infty}^{+\infty} p(z=0) \frac{\partial h}{\partial x} \, dx = -D,
\label{momsurf}
\end{equation}
where $p$ is the pressure perturbation associated with the waves and $h(x)$ is the terrain elevation. $D$ is the total drag exerted by the atmosphere on the orography. This can be viewed as an expression of Newton's third law.

It is clear from Equation (\ref{tendency}), that the vertical profile of $\rho \langle uw \rangle$, or equivalently of $M$, is crucial to define the drag exerted on the atmosphere by orographic gravity waves. However, difficulties arise when one attempts to evaluate $M$ for trapped lee waves. Since the wave solutions are most conveniently expressed in Fourier space, one might think that a way to evaluate $M$ would be by using Parseval's theorem (\citealp{Bretherton_1969}; \citealp{Teixeira_Miranda_2009}; \citealp{Nappo_2012}), which for horizontally bounded waves yields from Equation (\ref{momflux})
\begin{equation}
M=2 \pi i \rho \int_{-\infty}^{+\infty} \hat{u}^* \hat{w} \, dk,
\label{parseval}
\end{equation}
where $k$ is the horizontal wavenumber, $\hat{u}$ and $\hat{w}$ are the 1D Fourier transforms of $u$ and $w$, the asterisk denotes complex conjugate and $i=\sqrt{-1}$. Unfortunately, Equation (\ref{parseval}) cannot be used for inviscid trapped lee waves, at least when $z > 0$, because $u$ and $w$ do not approach zero downstream of the mountain as $x \rightarrow +\infty$. The momentum flux profile must therefore be obtained from an independent constraint, which generalizes Eliassen-Palm's theorem (\citealp{Eliassen_Palm_1960}; \citealp{Broad_2002}). This constraint can be derived in two alternative ways: either from direct manipulation of the equations of motion, or as a consequence of the conservation of wave activity under steady conditions (\citealp{Shepherd_1990}; \citealp{Lott_1998}). These two results will emerge as special cases of the more general treatment, including friction, to be presented next.

Consider the steady, linearized equations of motion for adiabatic 2D flow with the Boussinesq approximation (cf. \citealp{Teixeira_etal_2012}):
\begin{align}
U \frac{\partial u}{\partial x}+w \frac{dU}{dz} &= -\frac{1}{\rho_0} \frac{\partial p}{\partial x} -\lambda u, \label{eqmomhor} \\
U \frac{\partial w}{\partial x} &= -\frac{1}{\rho_0} \frac{\partial p}{\partial z} + b -\lambda w, \label{eqmomver} \\
U \frac{\partial b}{\partial x} + N^2 w &= 0, \label{eqbuoy} \\
\frac{\partial u}{\partial x} + \frac{\partial w}{\partial z} &= 0,
\label{eqmass}
\end{align}
where $b$ is the buoyancy perturbation associated with the waves, $N^2$ is the static stability of the mean incoming flow, and $\rho_0$ is a reference density (assumed to be constant). In this equation set, Rayleigh friction, with a (constant) damping coefficient $\lambda$, has been introduced only in the momentum balance equations, for simplicity.

If Equation (\ref{eqmomhor}) is multiplied by $u$ and Equation (\ref{eqmomver}) is multiplied by $w$ and both equations are added, this yields
\begin{equation}
U \frac{\partial}{\partial x} \left( \frac{u^2 + w^2}{2} \right)+uw \frac{dU}{dz} + \frac{\partial}{\partial x} \left( \frac{p u}{\rho_0} \right) + \frac{\partial}{\partial z} \left( \frac{p w}{\rho_0} \right) + N^2 \zeta w + \lambda \left( u^2 + w^2 \right)=0,
\label{interim1}
\end{equation}
where $\zeta$ is the vertical displacement of isentropes (or streamlines), which satisfies $w = U \partial \zeta/\partial x$, Equation (\ref{eqmass}) has been used, and a version of Equation (\ref{eqbuoy}) integrated with respect to $x$, yielding $b=-N^2 \zeta$, has also been used. Equation (\ref{eqmomhor}) may also be integrated with respect to $x$, yielding
\begin{equation}
U u + U \frac{dU}{dz} \zeta + \frac{p}{\rho_0}  + \lambda \int^x u \, dx = 0.
\label{interim2}
\end{equation}
This equation may be multiplied by $u$ or $w$, and differentiated with respect to $x$ or $z$, respectively, to eliminate the pressure terms in Equation (\ref{interim1}). When this is done, some terms cancel out, and the following equation is obtained:
\begin{equation}
U \frac{\partial}{\partial x} \left\{ \frac{w^2 - u^2}{2} + \frac{1}{2} \left( N^2 -U \frac{d^2 U}{dz^2} \right) \zeta^2 \right\} - U \frac{\partial}{\partial z} \left( u w \right) + \lambda w \int^x \left( \frac{\partial u}{\partial z} - \frac{\partial w}{\partial x} \right) dx = 0.
\label{constraint}
\end{equation}
When friction is neglected, and Equation (\ref{constraint}) is multiplied by $-\rho_0/U$, it takes the simpler form
\begin{equation}
-\frac{\partial}{\partial x} \left\{ \rho_0 \frac{w^2 - u^2}{2} + \frac{1}{2} \rho_0 \left( N^2 - U \frac{d^2 U}{dz^2} \right) \zeta^2 \right\} + \frac{\partial}{\partial z} \left( \rho_0  u w \right) = 0.
\label{inviscid}
\end{equation}
If this equation is integrated between $x=-\infty$ and a generic $x>0$ (assuming that any existing orography is centred at $x=0$), the following results:
\begin{equation}
\frac{\partial}{\partial z} \left( \rho_0 \int_{-\infty}^x u w \, dx \right) = \left\{ \rho_0 \frac{w^2 - u^2}{2} + \frac{1}{2} \rho_0 \left( N^2 - U\frac{d^2 U}{dz^2} \right) \zeta^2 \right\}(x),
\label{broadeq}
\end{equation}
where the fact that no wave perturbations exist as $x \rightarrow -\infty$ (i.e. upstream of the mountain) has been used. If $d^2U/dz^2$ is neglected, Equation (\ref{broadeq}) can be shown to be equivalent to Equation (10) of \citet{Broad_2002}. \citet{Broad_2002} then chose to focus on a value of $x$ corresponding to a phase of the trapped lee waves where both $u$ and $\zeta$ are zero (note that $u$ and $\zeta$ are in phase because $u = -(\partial/\partial z)(U \zeta)$, from $w=U \partial \zeta/\partial x$ and mass conservation, Equation (\ref{eqmass})). With these simplifications, Equation (\ref{broadeq}) reduces to Equation (15) of \citet{Broad_2002}. It is not obvious, however, why this phase of the trapped lee wave should be privileged. Clearly, there is no unique limit for Equation (\ref{broadeq}) when $x \rightarrow +\infty$, so according to Equation (\ref{momflux}) $M$ is mathematically ill-defined.

Another way to arrive at Equation (\ref{broadeq}) is using the wave activity balance equation (\citealp{Shepherd_1990}; \citealp{Lott_1998}). For a steady flow, the wave activity balance for 2D gravity waves with the Boussinesq approximation may be written
\begin{equation}
\frac{\partial F_x}{\partial x} + \frac{\partial F_z}{\partial z} = 0,
\label{actionbal}
\end{equation}
where
\begin{equation}
F_x = \frac{\rho_0 U}{N^2} b \left( \frac{\partial u}{\partial z} - \frac{\partial w}{\partial x} \right) + \frac{1}{2} \frac{\rho_0}{N^4}\left(N^2 - U \frac{d^2 U}{dz^2} \right) b^2 - \rho_0 \left( \frac{w^2 - u^2}{2} \right), \quad \quad F_z = \rho_0 u w
\label{pseudomom}
\end{equation}
(cf. Equations (23)-(25) of \citet{Lott_1998}, or the left-hand side of Equation (10) of \citet{Soufflet_etal_2022}). Equation (\ref{actionbal}) expresses the fact that the divergence of the (2D)  pseudo-momentum vector $(F_x, F_z)$ is zero. Using the equalities $b=-N^2 \zeta$, $w=U \partial \zeta/\partial x$ and $u= - (\partial/\partial z) (U \zeta)$, $F_x$ may be expressed as
\begin{equation}
F_x = \rho_0 U^2 \zeta \left( \frac{\partial^2 \zeta}{\partial x^2} + \frac{\partial^2 \zeta}{\partial z^2} + \frac{2}{U} \frac{dU}{dz} \frac{\partial \zeta}{\partial z} + \frac{N^2}{U^2} \zeta \right) -\frac{1}{2} \rho_0 \left\{ w^2 - u^2 + \left( N^2 - U \frac{d^2 U}{dz^2} \right) \zeta^2 \right\}.
\label{pseudox}
\end{equation}
The expression between the brackets on the left in this equation is zero (from the equation governing the behaviour of linear gravity waves, see \citet{Lin_2007}, Equation (5.3.1)).
This shows that Equation (\ref{actionbal}) is actually equivalent to Equation (\ref{inviscid}), that is, the first term between brackets in Equation (\ref{inviscid}) is minus a simplified form of the $x$ component of the pseudo-momentum vector. As far as the authors are aware, it is the first time that this is pointed out.

\begin{figure}[h!]
\begin{center}
\includegraphics[width=0.5\columnwidth]{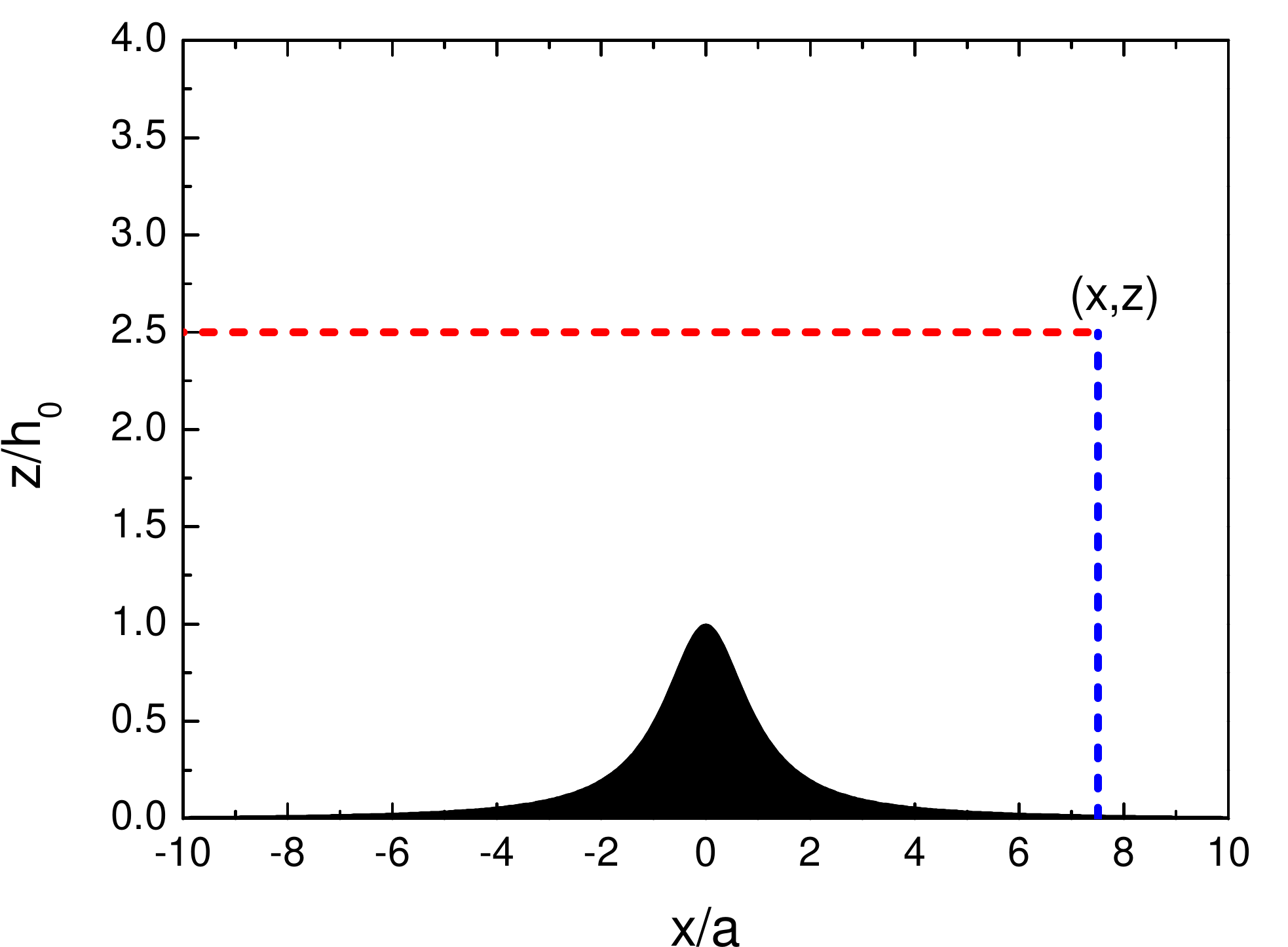}
\caption{Diagram showing the isolated mountain (with height $h_0$ and half-width $a$) that generates the trapped lee waves (in black) and the lines along which the momentum fluxes are integrated to obtain $P_z$ (red dashed line) and $P_x$ (blue dashed line) according to Equation (\ref{defspxpz}). $(x,z)$ is a generic point defining the upper limits of integration of $P_x$ and $P_z$. The mean flow is assumed to come from the left.}
\label{fig1}
\end{center}
\end{figure}

\citet{Lott_1998} noted that if Equation (\ref{actionbal}) (or Equation (\ref{inviscid})) is integrated horizontally between $-\infty$ and a generic $x$ (for $x$ downstream of the mountain) and vertically between 0 and $z>0$ (this is equivalent to integrating Equation (\ref{broadeq}) in the vertical between 0 and $z$), the following is obtained:
\begin{equation}
\rho_0 \int_{-\infty}^x u w \, dx - \frac{1}{2} \rho_0 \int_0^z \left\{ w^2 - u^2 + \left( N^2 - U \frac{d^2 U}{dz^2} \right) \zeta^2 \right\} dz = \rho_0 \int_{-\infty}^x u w(z=0) \, dx,
\label{constraint2}
\end{equation}
or in a more compact form
\begin{equation}
P_x + P_z = P_z(z=0),
\label{sumconst}
\end{equation}
where
\begin{equation}
P_x = \int_0^z F_x \, dz, \quad \quad P_z = \int_{-\infty}^x F_z \, dx,
\label{defspxpz}
\end{equation}
as defined by \citet{Lott_1998} or \citet{Soufflet_etal_2022}. Equation (\ref{constraint2}) (or Equation (\ref{sumconst})) shows that the sum of the integrated pseudomomentum fluxes along $x$ and along $z$ adds up to a constant value, which is equal to the vertical flux of horizontal pseudomomentum (or momentum) at the surface (by Equation (\ref{momsurf}), this is additionally equal -- in value -- to the surface pressure drag).
Figure \ref{fig1} shows as the horizontal red dashed line the domain of integration of $P_z$ and as the vertical blue dashed line the domain of integration of $P_x$. Clearly, any momentum flux emanating from the mountain (in black) must cross one of these lines.
Equation (\ref{constraint2}) suggests that inviscid linear theory cannot tell anything useful about the way the wave momentum fluxes force the mean flow, as this would require a depletion of the total integrated wave pseudo-momentum flux. In order to make progress, it is necessary to go back to Equation (\ref{constraint}), which includes friction. Taking Equation (\ref{pseudox}) into account, Equation (\ref{constraint}) can be considered equivalent to Equation (10) of \citet{Soufflet_etal_2022}, with the difference that the representation of friction on the right-hand side is more simplified. \citet{Soufflet_etal_2022} represent friction as a vertical diffusion that affects also heat instead of a Rayleigh damping affecting only momentum.

If Equation (\ref{constraint}) is integrated horizontally between $-\infty$ and $+\infty$, the first term with the $x$ derivative now cancels out, because when friction is included the trapped lee waves decay to zero downstream over a longer or shorter distance. Hence this integration yields:
\begin{equation}
\frac{\partial}{\partial z} \int_{-\infty}^{+\infty} u w \, dx = -\lambda \int_{-\infty}^{+\infty} \zeta \left( \frac{\partial u}{\partial z} - \frac{\partial w}{\partial x} \right) \, dx,
\label{step1}
\end{equation}
where $w= U \partial \zeta/\partial x$ has been used, Equation (\ref{constraint}) has been divided by $U$, and the frictional term has been integrated by parts. To obtain the vertical momentum flux itself (which is the only one relevant in this problem, as the horizontal momentum flux decays to zero downstream with the waves), Equation (\ref{step1}) needs to be integrated in the vertical, but now between a generic $z$ and $+\infty$ (where the $uw$ associated with the trapped waves is zero). This yields:
\begin{equation}
\int_{-\infty}^{+\infty} uw \, dx = \lambda \int_z^{+\infty} \int_{-\infty}^{+\infty} \zeta \left( \frac{\partial u}{\partial z} - \frac{\partial w}{\partial x} \right) \, dx \, dz.
\label{step2}
\end{equation}
When friction is included, the wave equation used to simplify Equation (\ref{pseudox}), which can be derived from the original equation set (\ref{eqmomhor})-(\ref{eqmass}), becomes more complicated, having some additional terms involving $\lambda$. However, in the limit of weak friction, it still takes approximately the same form.
In this approximation, which amounts to neglecting any terms proportional to powers of $\lambda$ higher than 1, Equation (\ref{step2}) may also be written as
\begin{equation}
\int_{-\infty}^{+\infty} u w \, dx = - \lambda \int_z^{+\infty} \frac{1}{U} \left(N^2 - U \frac{d^2 U}{dz^2} \right) \int_{-\infty}^{+\infty}  \zeta^2 \, dx \, dz.
\label{final}
\end{equation}
Clearly, under the necessary condition for vertical wave propagation in the trapping layer $N^2 - U d^2U/dz^2>0$, which is required for wave trapping, the term on the right-hand side of Equation (\ref{final}) will be always negative, which means that the momentum flux will equally be negative, and approach zero at high levels. This makes sense, since the waves are trapped within a layer, and $u w(z=0)$ should be negative by Newton's 3rd law.
In their studies on the diurnal evolution of trapped lee waves, \citet{Xue_Giorgietta_2021} and \citet{Xue_etal_2022} assume that the total momentum flux associated with trapped lee waves is zero (a consequence of their Equations (7) and (9), respectively). This results from the fact that, in the absence of a theory including friction, \citet{Xue_Giorgietta_2021} and \citet{Xue_etal_2022} inconsistently apply the inviscid theory of \citet{Broad_2002} to numerical simulation results where the trapped lee waves decay. Equation (\ref{final}) corrects this inconsistency.
Note that, in the term on the right-hand side of Equation (\ref{final}),
both $N$ and $U$ may vary with height, so they were kept inside the vertical integral.
Also worthy of note is that this term involves integrals over the whole wave field (in the horizontal and in the vertical directions). Therefore, the form and variation of $uw$ with height is a global property of the wave field (in particular, horizontally, from its point of generation to its point of total dissipation). It is this vertical flux of horizontal momentum that forces the mean flow in the area over the trapped lee wave field. Although, from Equation (\ref{final}), the momentum flux apparently should depend on $\lambda$, it actually does not, at least for low values of that parameter, as will be shown next. This is not obvious from Equation (\ref{final}), but may be understood intuitively in a qualitative way. The smaller $\lambda$ is, the more slowly the wave field spanned by the horizontal integral in Equation (\ref{final}) is expected to decay with $x$, therefore yielding a larger integral. It is plausible that these two effects could potentially cancel.

Interestingly, for weak friction, Equation (\ref{momsurf}) remains approximately valid. This can be shown departing from Equation (\ref{interim2}). If Equation (\ref{interim2}) is multiplied by $w$ and integrated horizontally between $-\infty$ and $+\infty$, the result is the following:
\begin{equation}
U \int_{-\infty}^{+\infty} uw \, dx + \frac{1}{\rho_0} \int_{-\infty}^{+\infty} p w \, dx - \lambda U \int_{-\infty}^{+\infty} u \zeta  \, dx = 0,
\label{altern}
\end{equation}
where $w= U \partial \zeta/\partial x$ has been used, the term coming from the second term of Equation (\ref{interim2}) vanishes, because it can be expressed as the horizontal integral of $(1/2) U^2 (dU/dz) \partial (\zeta^2)/\partial x$, and the last term has been integrated by parts.
If Equation (\ref{altern}) is multiplied by $\rho_0/U$ and applied at $z=0$, it becomes
\begin{equation}
\rho_0 \int_{-\infty}^{+\infty} uw (z=0) \, dx + \int_{-\infty}^{+\infty} p(z=0) \frac{\partial h}{\partial x} \, dx - \lambda \rho_0 \int_{-\infty}^{+\infty} u h \, dx = 0,
\label{altern2}
\end{equation}
where $w(z=0)=\partial h/\partial x$ and $\zeta(z=0)=h$ have been used. Since $h$ is only non-zero near the mountain, the integral on the last term does not increase indefinitely as $\lambda$ decreases, and therefore the whole term vanishes as $\lambda \rightarrow 0$ (unlike the last term in Equation (\ref{final})). Therefore, the conclusion is that Equation (\ref{momsurf}) still holds, as intended.

\subsection{Application to Scorer's atmosphere}

Equation (\ref{final}) is the main outcome of the preceding section. It describes the variation of the vertical flux of horizontal momentum associated with the trapped lee waves under the linear approximation,
representing frictional effects as a Rayleigh damping. In order to proceed further, it is necessary to assume a specific atmospheric profile, which will determine the form of $\zeta$ inside the integral on the right-hand side of Equation (\ref{final}). A crucial question is how to specify $\zeta$ itself. Clearly, in order for the integrals on the right-hand side of Equation (\ref{final}) to converge, the wave field must be bounded (which is consistent with the existence of friction). Analytical expressions for $\zeta$ (in physical space) in trapped lee waves with friction do not exist, even under the linear approximation (\citealp{Smith_etal_2006}; \citealp{Jiang_etal_2006}). Asymptotic solutions for $\zeta$ (downstream of the mountain) from inviscid linear theory are analytical (\citealp{Scorer_1949} \citealp{Mitchell_etal_1990}), but they (accurately) extend indefinitely in space, so if they were used in Equation (\ref{final}) without adaptation, the horizontal integral on the right-hand side would not converge. Here, a compromise will be made, which can be shown to be increasingly accurate as $\lambda$ becomes smaller: the inviscid solutions for $\zeta$ will be used, but multiplied by an exponentially decaying factor that accounts for the effect of weak friction. Since, for the calculation of the right-hand side of Equation (\ref{final}), the primary effect of weak friction is to limit the wave field to a finite extent in space, but apart from this modulation the solution for $\zeta$ is virtually indistinguishable from the inviscid one, this seems a very reasonable approach.

More specifically, it will be assumed that
\begin{equation}
\zeta = \zeta_I {\rm e}^{-k_I x},
\label{defzeta}
\end{equation}
where $\zeta_I$ is the inviscid form of $\zeta$ and $k_I$ is the imaginary part of the wavenumber associated with a spectral representation of $\zeta$. While for an inviscid solution $k_I=0$, in the solution with friction, $k=K_R + i k_I$. For all purposes, in what follows it will be assumed that $k=k_R$, and $k_I$ will be assumed to be non-zero, but very small, in Equation (\ref{defzeta}) via a definition to be presented, relating $k_I$ to $\lambda$. For the time being, it is sufficient to recognize that Equation (\ref{defzeta}) is accurate. Inserting Equation (\ref{defzeta}) into Equation (\ref{final}) yields
\begin{equation}
\int_{-\infty}^{+\infty} u w \, dx = - \lambda \int_{0}^{+\infty} \int_z^{+\infty} U l^2  \zeta_I^2 \, dz \, {\rm e}^{-2 k_I x} \, dx,
\label{final2}
\end{equation}
where $l=[N^2/U^2 - (1/U)(d^2 U/dz^2)]^{1/2}$ is the Scorer parameter, for convenience the integrations over $x$ and $z$ have been swapped (note that the exponential term does not depend on $z$), and it has been noted that the trapped lee waves only exist downstream of the mountain (assumed to be centred at $x=0$), hence the lower limit of integration in $x$ has been changed from $-\infty$ to 0. All of this ensures that the horizontal integral on the right-hand side of Equation (\ref{final2}) converges.

  The inviscid solution $\zeta_I$ for the two-layer atmosphere of \citet{Scorer_1949} is easily obtained from the corresponding solutions for the Fourier transforms of flow variables in \citet{Teixeira_etal_2013a}, in the same way as this was done in \citet{Teixeira_Miranda_2017} for similar, but 3D waves. The solution corresponds to a monochromatic wave resulting from a singularity in the Fourier transform, as originally noted by \citet{Scorer_1949} (the same singularity that is responsible for the drag from trapped lee waves in \citet{Teixeira_etal_2013a}), and can be written
\begin{align}
\zeta_{I1} &= -4 \pi \frac{ \hat{h}(k_L) m_1(k_L) n_2(k_L) \sin \{m_1(k_L) z\}}{k_L \left\{ 1 + n_2(k_L) H \right\}} \sin(k_L x) \quad {\rm if} \quad 0< z< H, \nonumber \\
\zeta_{I2} &= -\frac{4 \pi}{(l_1^2 - l_2^2)^{1/2}} \frac{\hat{h}(k_L) m_1^2(k_L) n_2(k_L) {\rm e}^{-n_2(k_L) (z-H)}}{k_L \left\{ 1 + n_2(k_L) H \right\}} \sin(k_L x) \quad {\rm if } \quad z>H,
\label{leezeta}
\end{align}
where $\zeta_{I1}$ is the solution in the lower layer, $\zeta_{I2}$ is the solution in the upper layer, $H$ is the height of the interface between the two layers and $\hat{h}$ is the Fourier transform of the terrain elevation $h$. $k_L$ is the horizontal wavenumber of the resonant trapped lee wave mode, $m_1=(l_1^2 - k^2)^{1/2}$ is the vertical wavenumber in the lower layer, and $n_2=(k^2 - l_2^2)^{1/2}$ is the vertical spatial decay rate of the waves in the upper layer, where they are evanescent. $l_1=N_1/U$ and $l_2=N_2/U$ are the Scorer parameters in the lower and upper layers, respectively, where $N_1$ and $N_2 < N_1$ are the corresponding Brunt-V\"ais\"al\"a frequencies (since in Scorer's atmosphere $d^2 U/dz^2=0$). Note that, unlike in \citet{Teixeira_etal_2013a} or \citet{Teixeira_Miranda_2017}, a sum is not included in Equation (\ref{leezeta}) because the results will only focus on a single trapped lee wave mode (the lowest one), for simplicity. But, to be strictly correct, the sum over all wave modes should be included, as in \citet{Teixeira_etal_2013a}.
Strictly speaking, the asymptotic solution Equation (\ref{leezeta}) is only accurate some distance downstream of the mountain ($x>0$), as noted by \citet{Scorer_1949}. But both use of this approximation in the calculation of the horizontal integral on the right-hand side of Equation (\ref{final2}) and the adoption of the lower limit of integration 0 (centred on the mountain), while subject to some errors for finite friction, are actually extremely accurate for weak friction. This is because, for weak enough friction, the portion of the wave field that is far downstream of the mountain gives an overwhelmingly dominant contribution to the integral, whereas the contribution of the wave field near to the mountain, as well as the exact value of the lower limit of integration, are essentially irrelevant. 

The integral in $z$ in Equation (\ref{final2}) is calculated first. The result can be shown to be
\begin{align}
  \int_z^{+\infty} U l_1^2 \zeta_I^2 \, dz &= \frac{8 \pi^2 U}{l_1^2 - l_2^2} \frac{\left| \hat{h}(k_L) \right|^2 m_1^2(k_L) n_2(k_L)}{k_L^2 \left\{ 1 + n_2(k_L) H \right\}^2} \left\{ l_2^2 m_1^2(k_L) + l_1^2 (l_1^2 - l_2^2) n_2(k_L) \left( H - z \right. \right. \nonumber \\
&\left. \left. + \frac{1}{2 m_1(k_L)} \left[ \sin\{2 m_1(k_L) z\} - \sin \{2 m_1(k_L) H\} \right] \right) \right\} \sin^2 (k_L x) \quad {\rm if} \quad z<H, \nonumber \\
\int_z^{+\infty} U l_2^2 \zeta_I^2 \, dz &= \frac{8 \pi^2 U l_2^2}{l_1^2 - l_2^2} \frac{\left| \hat{h}(k_L) \right|^2 m_1^4(k_L) n_2(k_L)}{k_L^2 \left\{ 1 + n_2(k_L) H \right\}^2} {\rm e}^{-2 n_2(k_L) (z-H)} \sin^2 (k_L x) \quad {\rm if} \quad z>H.
\label{defints}
\end{align}
If Equation (\ref{defints}) is inserted into Equation (\ref{final2}), only the factors $\sin^2 (k_L x)$ depend on $x$, hence the following integral will arise:
\begin{equation}
\int_0^{+\infty} \sin^2 (k_L x) {\rm e}^{-2 k_I x} \, dx \approx \frac{1}{4 k_I},
\label{intsimp}
\end{equation}
where the approximation in Equation (\ref{intsimp}) becomes progressively more accurate as $k_I \rightarrow 0$. When this result is used, Equation (\ref{final2}) becomes
\begin{align}
\int_{-\infty}^{+\infty} uw \, dx &= - \frac{2 \pi^2 \lambda U}{k_I (l_1^2 - l_2^2)} \frac{\left| \hat{h}(k_L) \right|^2 m_1^2(k_L) n_2(k_L)}{k_L^2 \left\{ 1 + n_2(k_L) H \right\}^2} \left\{ l_2^2 m_1^2(k_L) + l_1^2 (l_1^2 - l_2^2) n_2(k_L) \left( H - z \right. \right. \nonumber \\
&\left. \left. + \frac{1}{2 m_1(k_L)} \left[ \sin\{2 m_1(k_L) z\} - \sin \{2 m_1(k_L) H \} \right] \right) \right\} \quad {\rm if} \quad z<H, \nonumber \\ 
\int_{-\infty}^{+\infty} uw \, dx &= - \frac{2 \pi^2 \lambda U l_2^2}{k_I (l_1^2 - l_2^2)} \frac{\left| \hat{h}(k_L) \right|^2 m_1^4(k_L) n_2(k_L)}{k_L^2 \left\{ 1 + n_2(k_L) H \right\}^2} {\rm e}^{-2 n_2(k_L) (z-H)} \quad {\rm if} \quad z>H.
\label{final3}
\end{align}

It remains to evaluate $k_I$. One might naively consider assuming that $k_I= \lambda/U$, given the form of the Rayleigh damping terms in Equations (\ref{eqmomhor})-(\ref{eqmomver}), but this is not correct. In order to obtain an accurate definition for $k_I$, it is necessary to go back to the wave solutions. In the inviscid trapped lee wave solution, the wavelength of the wave is determined by the (real) wavenumber at which the Fourier transform of the solution has a singularity. When friction is added to the problem, this `singularity' (which can no longer be classified as such in the usual sense) corresponds to a complex value of the wavenumber at which the Fourier transform becomes infinite. The imaginary part of that wavenumber is $k_I$. The relevant wave solutions can be found in the Appendix of \citet{Teixeira_etal_2013a}. For example, from Equation (A1) shown there, it can be seen that the Fourier transform of the vertical velocity (which is proportional to the coefficient $a_1$) becomes singular (i.e. infinite) if the denominator of $a_1$ is zero, that is if
\begin{equation}
m_1 \cos(m_1 H) - i m_2 \sin(m_1 H) = 0.
\label{singular}
\end{equation}
In this equation, $m_1$, $m_2$, and the corresponding horizontal wavenumber $k$ for which Equation (\ref{singular}) is satisfied may all be complex. To determine $k_I$, it must be noted that $m_1 = m_{1R} + i m_{1I}$, $m_2 = m_{2R} + i m_{2I}$ and $k = k_R +i k_I$. This encompasses, for example, the cases in which $m_2$ is purely imaginary (in which case $m_{2I}$ is named $n_2$ -- cf. \citet{Teixeira_etal_2013a}). In order for Equation (\ref{singular}) to be usable, it is necessary to express the sine and cosine functions in complex form, and expand all the variables into their real and imaginary parts. Since the aim is to take friction into account, it is also necessary to assume definitions for $m_1$ and $m_2$ that are consistent with Equations (\ref{eqmomhor})-(\ref{eqmass}). This is provided by Equation (12) of \citet{Teixeira_etal_2012}, which is reproduced next:
\begin{equation}
m_j^2 = \frac{l_j^2}{1 - i \frac{\lambda}{U k}} - k^2,
\label{vertwav}
\end{equation}
where $j=1,2$ depending on whether it refers to the lower or upper layer, respectively. The deceptively simple form of Equation (\ref{vertwav}), where the effect of friction is contained in $\lambda$, conceals the fact that both $m_j$ and $k$ are complex, yielding much lengthier expressions for the real and imaginary parts of this equation (of which an example, for real $k$, is provided by Equations (14)-(15) of \citet{Teixeira_etal_2013a}). To obtain $k_I$, both the real and imaginary parts of Equation (\ref{singular}) must be satisfied, which in a general case would produce equations that are too complicated. However, for weak friction, some simplifications are possible. For example, it is known that $\lambda$ is small, but $k_I$ is also expected to be small, since it is zero in the inviscid approximation. Additionally $m_{1I}$ and $m_{2R}$ are also expected to be small (they are also zero in the inviscid approximation). With these assumptions, only the leading order terms are not neglected in the equations for the real and imaginary parts of Equation (\ref{singular}). After a substantial amount of algebra, it turns out that, to leading order, Equation (\ref{singular}) reduces to
\begin{align}
&m_{1R} + m_{2I} \tan(m_{1R} H) = 0, \label{singular21} \\
&m_{1I} \left( 1 + m_{2I} H \right) - \left( m_{2R} + m_{1R} m_{1I} H \right) \tan(m_{1R} H) = 0,  
\label{singular22}
\end{align}
and Equation (\ref{vertwav}) can be expressed as
\begin{align}
  m_{jR}^2 &= l_j^2 - k_R^2, \label{eqsform1} \\
  2 m_{jR} m_{jI} &= l_j^2 \frac{\lambda}{U k_R} - 2 k_R k_I,
\label{eqsform2}
\end{align}
where $j=1,2$ apply to the lower or upper layer, respectively. Note that Equations (\ref{singular21}) and (\ref{eqsform1}) define the wave resonance condition and the vertical wavenumber in the same way as in purely inviscid theory (with $m_{jR}=m_j$, $m_{2I}=n_2$ and $k_R=k$), whereas Equations (\ref{singular22}) and (\ref{eqsform2}) are equations where each term is of first order in the small quantities mentioned above. $m_{1I}$ is small in the lower layer, whereas $m_{2R}$ is small in the upper layer, and both $\lambda$ and $k_I$ are small in both layers. From Equations (\ref{singular21})-(\ref{eqsform2}), it is possible to obtain $k_I$ in terms of $k_R$. The final result is:
\begin{equation}
k_I= \frac{\lambda}{2 U} \frac{k_R^2 + l_1^2 n_2 H}{k_R^2 (1 + n_2 H)},
\label{defki}
\end{equation}
where $m_{2I}=n_2$ has been used.
Noting that, for a flow with weak friction that satisfies Equation (\ref{singular}), $k_R=k_L$, and using Equation (\ref{defki}) with $n_2=n_2(k_L)$ in Equation (\ref{final3}), the following expressions for the momentum flux are finally obtained:
\begin{align}
M&= \rho_0 \int_{-\infty}^{+\infty} uw \, dx = - \frac{4 \pi^2 \rho_0 U^2}{l_1^2 - l_2^2}  \frac{\left| \hat{h}(k_L) \right|^2 m_1^2(k_L) n_2(k_L)}{\left\{ 1 + n_2(k_L) H \right\} \left\{ k_L^2 + l_1^2 n_2(k_L) H \right\}} \left\{ l_2^2 m_1^2(k_L) + l_1^2 (l_1^2 - l_2^2) n_2(k_L) \left( H - z \right. \right. \nonumber \\
&\left. \left. + \frac{1}{2 m_1(k_L)} \left[ \sin\{2 m_1(k_L) z\} - \sin \{2 m_1(k_L) H\} \right] \right) \right\} \quad {\rm if} \quad z<H, \nonumber \\ 
M&= \rho_0 \int_{-\infty}^{+\infty} uw \, dx = - \frac{4 \pi^2 \rho_0 U^2 l_2^2}{l_1^2 - l_2^2}  \frac{\left| \hat{h}(k_L) \right|^2 m_1^4(k_L) n_2(k_L) \, {\rm e}^{-2 n_2(k_L) (z-H)}}{ \left\{ 1 + n_2(k_L) H \right\} \left\{ k_L^2 + l_1^2 n_2(k_L) H \right\}}  \quad {\rm if} \quad z>H.
\label{final4}
\end{align}
From Equation (\ref{final4}), it can be concluded not only that the momentum flux is continuous at $z=H$ (as it should), but also that it reduces at the surface to minus the total pressure drag (confirming Equation (\ref{altern2})), namely:
\begin{equation}
M(z=0)= - 4 \pi^2 \rho_0 U^2 \left| \hat{h}(k_L) \right|^2 \frac{m_1^2(k_L) n_2(k_L)}{1 + n_2(k_L) H},
\label{mindrag}
\end{equation}
which should be compared with Equation (25) of \citet{Teixeira_etal_2013a}. However, perhaps the most important feature of (\ref{final4}) is that the momentum flux is independent of $\lambda$, because $\lambda$ in the numerator of the fraction in Equation (\ref{final3}) cancels out with $k_I$ (which is proportional to $\lambda$ according to Equation (\ref{defki})) in the denominator of the same fraction. This feature quantifies the intuitive qualitative result that was  mentioned before when discussing Equation (\ref{final}), about the inverse variation of the spatial extent of the trapped lee wave train with $\lambda$.

Equation (\ref{final4}) is the main result of the present section. It gives a closed-form expression (except for the necessarily numerical root-finding procedure to determine $k_L$) for the momentum flux associated with trapped lee waves in the two-layer atmosphere of \citet{Scorer_1949}. It probably represents the closest one can get to an inviscid solution for the momentum flux produced by trapped lee waves for that model atmosphere, but, as we saw, the inclusion of friction (no matter how weak), is essential to obtain it consistently. Hereafter, Equation (\ref{final4}) will be called the quasi-inviscid theory or solution.
For comparison, the purely inviscid solution for $M$ (with an upper limit of integration $x$ replacing $+\infty$) from \citet{Broad_2002} (deriving from the first term on the right-hand side of Equation (\ref{broadeq}), involving $w^2$) takes the form, for the two layer atmosphere of \citet{Scorer_1949},
\begin{align}
M_{B}&= - \frac{4 \pi^2 \rho_0 U^2}{l_1^2 - l_2^2}  \frac{\left| \hat{h}(k_L) \right|^2 m_1^2(k_L) n_2(k_L)}{\left\{ 1 + n_2(k_L) H \right\}^2} \left\{ m_1^2(k_L) + (l_1^2 - l_2^2) n_2(k_L) \left( H - z \right. \right. \nonumber \\
&\left. \left. + \frac{1}{2 m_1(k_L)} \left[ \sin\{2 m_1(k_L) z\} - \sin\{2 m_1(k_L) H\} \right] \right) \right\} \cos^2(k_L x) \quad {\rm if} \quad z<H, \nonumber \\ 
M_{B}&= - \frac{4 \pi^2 \rho_0 U^2}{l_1^2 - l_2^2}  \frac{\left| \hat{h}(k_L) \right|^2 m_1^4(k_L) n_2(k_L) \, {\rm e}^{-2 n_2(k_L) (z-H)}}{ \left\{ 1 + n_2(k_L) H \right\}^2}  \cos^2(k_L x) \quad {\rm if} \quad z>H.
\label{broadmom}
\end{align}
The momentum flux deriving from the second and third terms on the right-hand side of Equation (\ref{broadeq}) (involving $u^2$ and $\zeta^2$), on the other hand, may be written
\begin{align}
M_{NB}&= -\frac{4 \pi^2 \rho_0 U^2}{l_1^2 - l_2^2}  \frac{\left| \hat{h}(k_L) \right|^2 m_1^2(k_L) n_2(k_L)}{k_L^2 \left\{ 1 + n_2(k_L) H \right\}^2} \left( m_1^2(k_L) \{l_2^2- n_2^2(k_L)\} + k_L^2 (l_1^2 - l_2^2) n_2(k_L) (H - z) \right. \nonumber \\
&\left. + \frac{\{l_1^2 + m_1^2(k_L)\} (l_1^2 - l_2^2) n_2(k_L)}{2 m_1(k_L)} \left[ \sin \{2 m_1(k_L) z \} - \sin \{2 m_1(k_L) H \} \right] \right) \sin^2(k_L x) \quad {\rm if} \quad z<H, \nonumber \\ 
M_{NB}&= -\frac{4 \pi^2 \rho_0 U^2}{l_1^2 - l_2^2}  \frac{\left| \hat{h}(k_L) \right|^2 m_1^4(k_L) n_2(k_L) \left\{ l_2^2-n_2^2(k_L) \right\} \, {\rm e}^{-2 n_2(k_L) (z-H)}}{k_L^2 \left\{ 1 + n_2(k_L) H \right\}^2}  \sin^2(k_L x) \quad {\rm if} \quad z>H.
\label{nonbroad}
\end{align}
The total inviscid solution is the sum of Equations (\ref{broadmom}) and (\ref{nonbroad}), $M_B+M_{NB}$. Note that, unlike in Equation (\ref{final4}), $M_B$ and $M_{NB}$ depend on the upper limit of integration $x$ via the phase of the (infinite) trapped lee wave (the $\cos^2(k_L x)$ and $\sin^2(k_L x)$ factors included in these expressions). Additionally, $M_B$ and $M_{NB}$ are in quadrature. So, when $\sin(k_L x)=0$ (the situation envisaged by \citet{Broad_2002}) $M_{NB}=0$ and $M_{B}$ is a maximum, whereas when $\cos(k_L x)=0$, $M_B=0$ and $M_{NB}$ is a maximum. The form of the variation with height of $M_{NB}$ is, however, very different from that of $M_B$ (or of $M$ as given by Equation (\ref{final4})), as will be seen next.

\section{Numerical models}

To compare and validate the previous results from linear theory, two models will be used. One of them is a linear model that includes friction exactly in the same form as envisaged in the theory described above, but where the Rayleigh damping coefficient may take an arbitrary value. The second model is a numerical model where the fully nonlinear dynamics of the waves can be represented. These are described next in turn.

\subsection{Linear model with friction}

This linear model departs from exactly the same equation set as used in the preceding calculations, comprising Equations (\ref{eqmomhor})-(\ref{eqmass}). The approach is similar to that adopted in \citet{Teixeira_etal_2012}. Since friction is always non-zero, and hence the flow perturbations associated with the trapped lee waves always decay downstream (i.e. they are bounded spatially), these perturbations can be expressed as Fourier integrals. The Fourier transform of the vertical velocity $\hat{w}$ (in terms of which all other flow perturbation variables may be expressed) satisfies Equation (6) of \citet{Teixeira_etal_2012}. The vertical wavenumber of the waves $m$ (which for the two-layer atmosphere of Scorer takes different values in the lower and upper layer) may be expressed as in Equation (12) of \citet{Teixeira_etal_2012} (with $l_0$ replaced by $l_1$ in the lower layer and by $l_2$ in the upper layer). This wavenumber is of course complex, with real and imaginary parts $m_R$ and $m_I$ given by Equations (14)-(15) of \citet{Teixeira_etal_2012} (again with $l_0$ replaced by $l_1$ or $l_2$). The procedure to obtain $m_R$ and $m_I$ is entirely analogous to that described in \citet{Teixeira_etal_2012}, with the difference that boundary conditions for $\hat{w}$ and $d\hat{w}/dz$ (resulting from continuity of pressure) must be satisfied at $z=H$, in the same way as was necessary for the derivation of Equation (\ref{leezeta}).

From Equation (\ref{parseval}), which in this case is valid because $w$ is bounded spatially, and from mass conservation, Equation (\ref{eqmass}), expressed in terms of Fourier transforms, it can be shown that the momentum flux
is given by
\begin{equation}
  M= 4 \pi \rho_0 \int_0^{+\infty} \left[ {\rm Im}(\hat{w}) \frac{\partial}{\partial z} \left\{ {\rm Re}(\hat{w}) \right\} - {\rm Re}(\hat{w}) \frac{\partial}{\partial z} \left\{ {\rm Im}(\hat{w}) \right\} \right] \frac{dk}{k},
  \label{momfric}
\end{equation}
where the fact that the integrand is symmetric with respect to $k$ has been taken into account. 
The integral in Equation (\ref{momfric}) is not analytical in the presence of friction, i.e. with $\lambda >0$, and so must be calculated numerically, using a Gauss-Legendre quadrature algorithm. In this calculation, $\lambda$ cannot be too low, otherwise the contribution to the integral concentrates progressively more around a singularity (corresponding to the inviscid resonant trapped lee wave mode), and the numerical integration procedure fails.
In practice, and as will be seen, a friction coefficient as small as necessary to make the results converge to those of the quasi-inviscid theory (presented previously), can be used.

\subsection{Nonlinear numerical model}

Nonlinear numerical simulations are carried out using the micro-to-mesoscale model FLEX (\citealp{Argain_2003}; \citealp{Argain_etal_2009}, \citeyear{Argain_etal_2017}). This is a 2D fully nonlinear numerical model using curvilinear orthogonal coordinates with grid refinement near the ground, which is able to accurately represent boundary layer flows. Here the model is run in inviscid mode, since the primary aim is to test the theoretical results presented above before any additional flow complications are considered.

The domain of integration consists of 556 grid points in the horizontal direction and 2244 grid points in the vertical. With a grid spacing of $180 \,{\rm m}$ in the horizontal and $7 \,{\rm m}$ in the vertical, this yields a domain size of $100 \,{\rm km}$ in the horizontal and $15.7 \,{\rm km}$ in the vertical. The time step of the simulations is $0.5\,{\rm s}$.
In all experiments, the orography corresponds to a 2D bell-shaped mountain, with terrain elevation given by
\begin{equation}
  h(x)=\frac{h_0}{1 + (x/a)^2},
  \label{bellshape}
\end{equation}
where the mountain half-width is $a=1 \,{\rm km}$ and the mountain height is $h_0=10 \,{\rm m}$.
The wind speed is constant (the lower boundary condition is free-slip) with a magnitude $U=10 \,{\rm m} \,{\rm s}^{-1}$, and the Brunt-V\"ais\"al\"a frequency in the lower layer is $N_1= 0.02 \,{\rm s}^{-1}$, yielding a Scorer parameter of $l_1=N_1/U=2 \times 10^{-3} \,{\rm m}^{-1}$. The Brunt-V\"ais\"al\"a frequency in the upper layer is defined by the assumed ratio $N_2/N_1=l_2/l_1$, which varies between simulations. The flow regime is strongly linear ($l_1 h_0=0.02$) and strongly non-hydrostatic ($l_1 a= 2$), which favours the existence of trapped lee waves that can dominate the flow \citep{Teixeira_etal_2013a}. 
Given the above parameters, the horizontal domain length corresponds to $100 a$, with $10 a$ extending upstream of the mountain and $90 a$ extending downstream. The purpose of this asymmetry is to focus on the trapped lee waves, which only exist downstream of the orography and can persist for a long distance, especially in inviscid numerical simulations. The vertical domain length corresponds to 5 vertical wavelengths of the hydrostatic mountain waves for the parameters of the lower layer, $\lambda_1=2 \pi U/N_1=2 \pi/l_1 \approx 3.1 \,{\rm km}$.
Sponge layers with thickness $4 a$
exist at the upstream and downstream boundaries of the domain.  A sponge layer with thickness $ 2 \lambda_1 \approx 6.3\,{\rm km}$ exists at the top of the domain.

The simulations are run until a time when an approximate steady state is reached by the wave momentum flux profile. This time can vary approximately between $100 a/U \approx 2.8$ hr and $400 a/U \approx 11.1$ hr.

\section{Preliminary results}

Preliminary tests to the theory developed in section \ref{method} will be divided into two parts. Firstly, comparisons will be made with perfectly inviscid solutions, of the same type as those produced by \citet{Broad_2002} and included in the treatment presented above. Secondly, the quasi-inviscid solutions (with vanishing but non-zero friction), which constitute the bulk of the preceding theoretical treatment and are those of greatest practical importance, will be tested.

\begin{figure}[h!]
\begin{center}
 (a)\includegraphics[width=0.45\columnwidth]{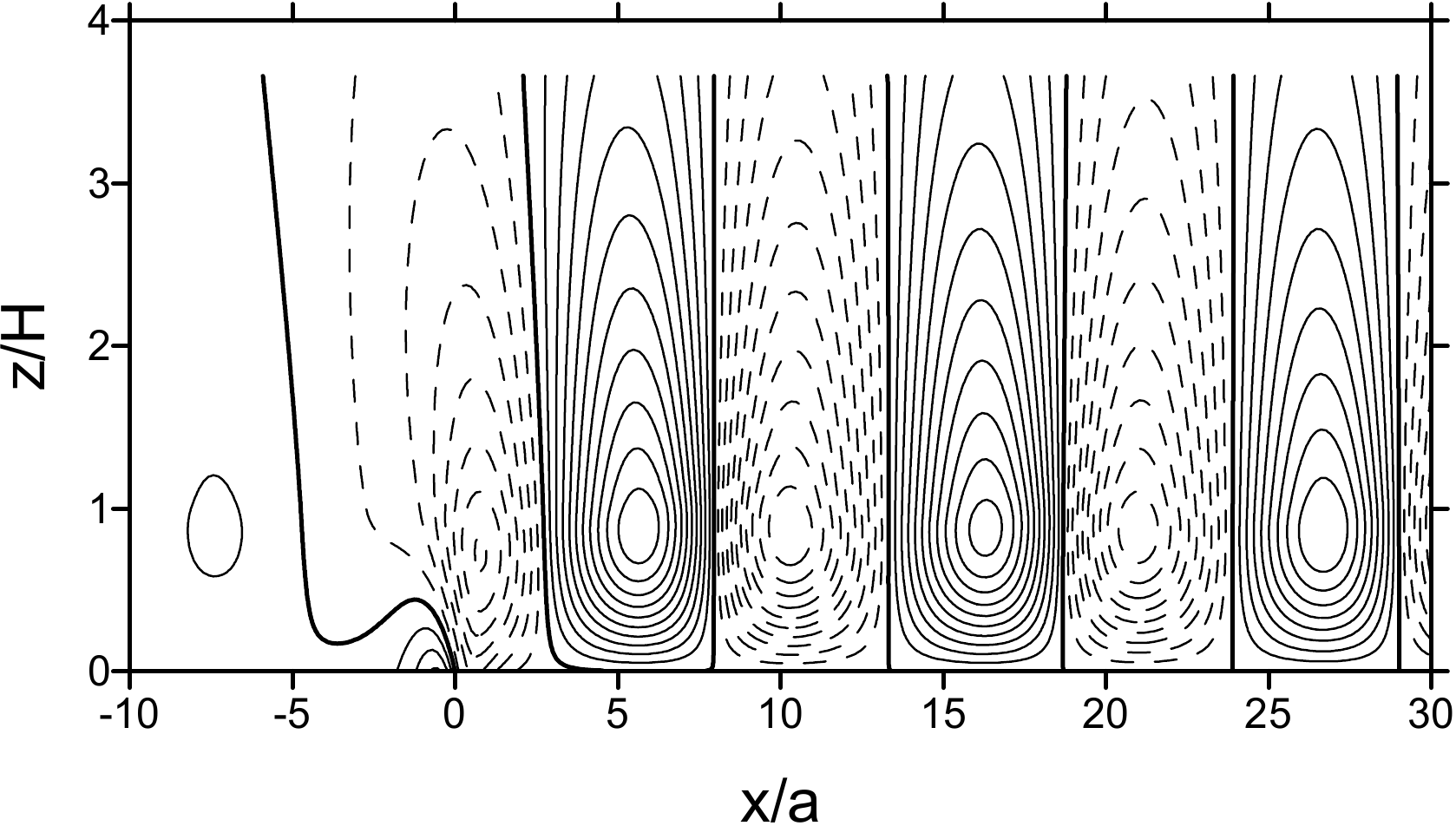}
 (b)\includegraphics[width=0.45\columnwidth]{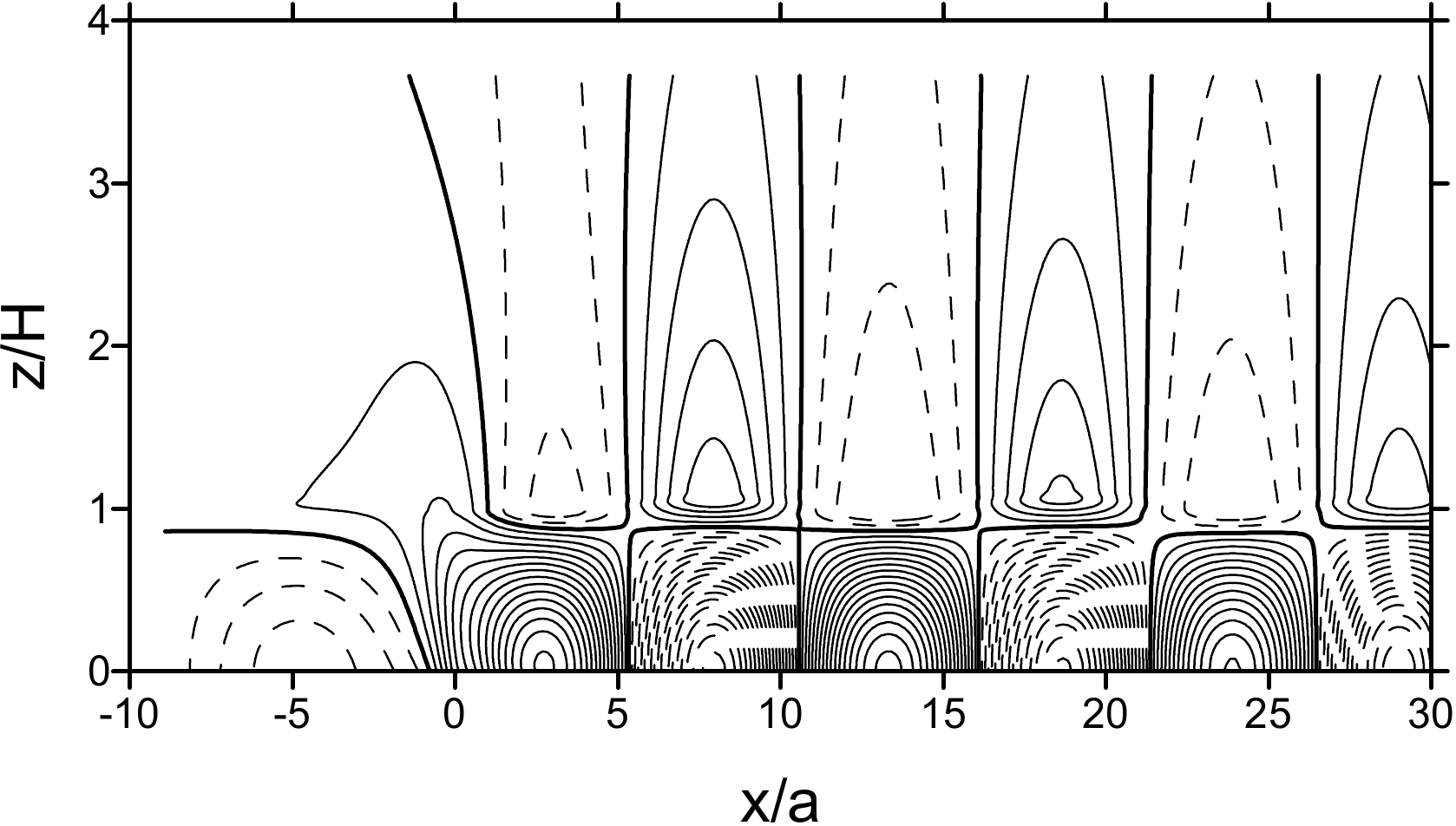} 
 (c)\includegraphics[width=0.45\columnwidth]{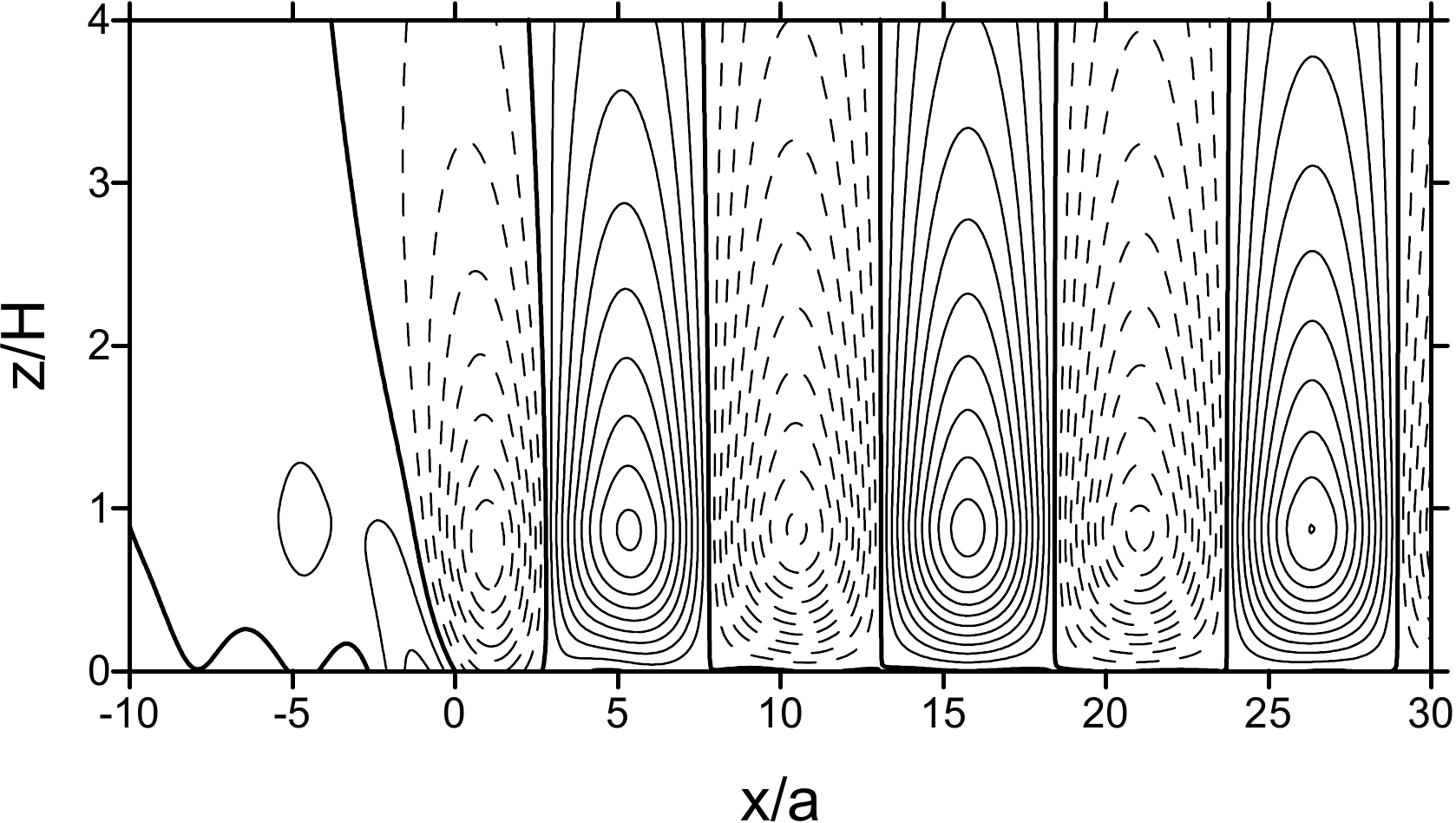}
 (d)\includegraphics[width=0.45\columnwidth]{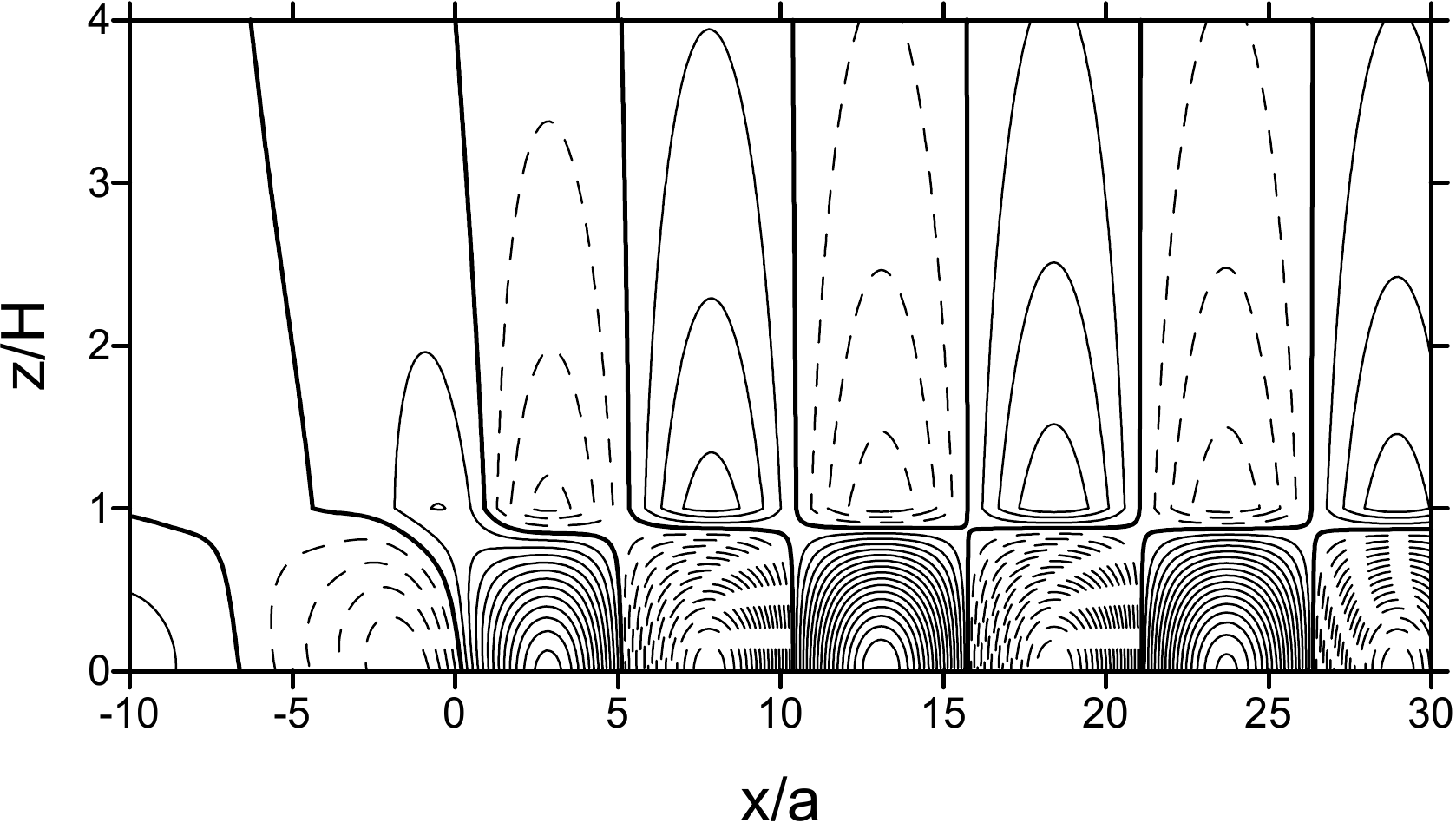}
 (e)\includegraphics[width=0.45\columnwidth]{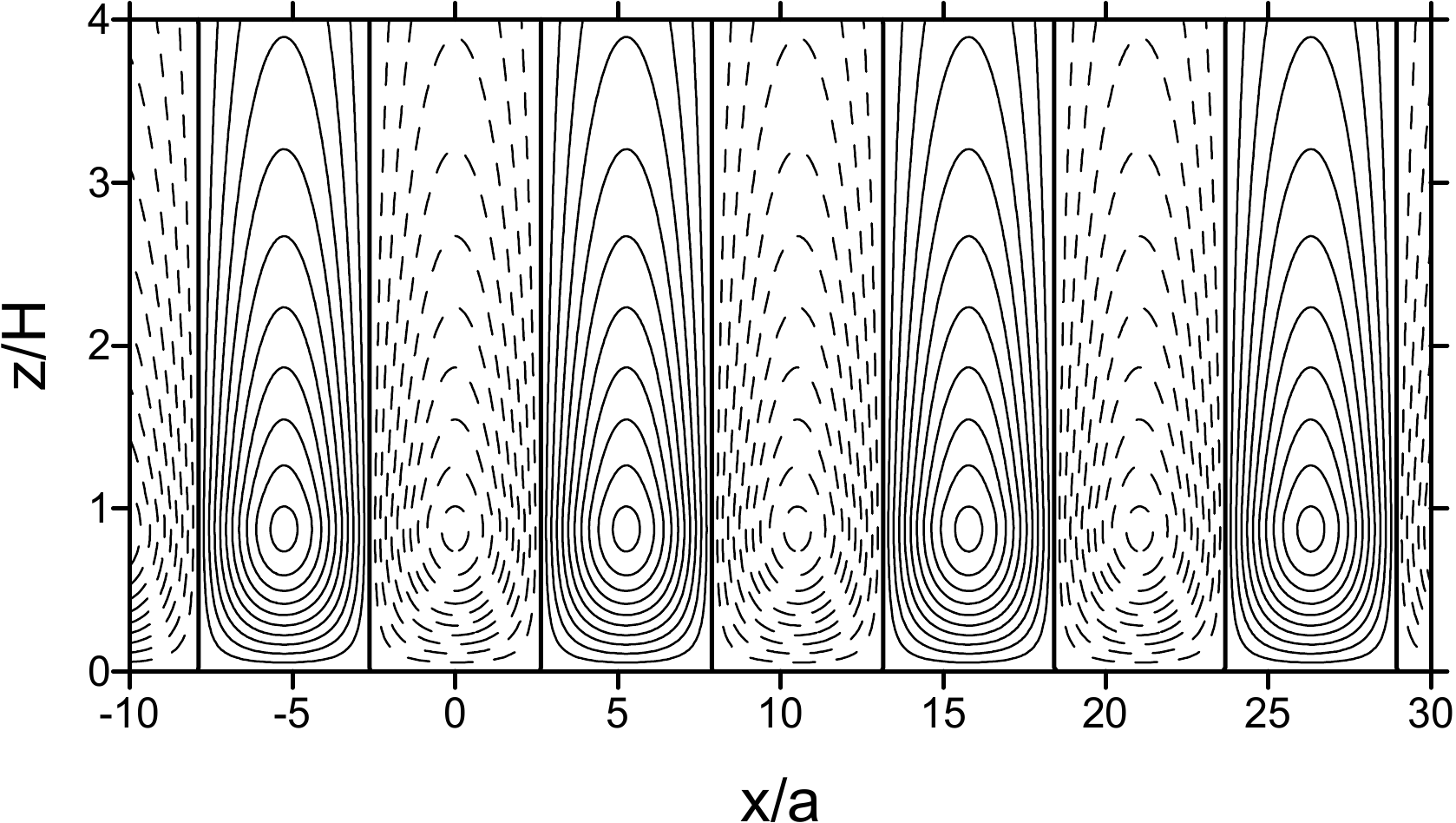}
 (f)\includegraphics[width=0.45\columnwidth]{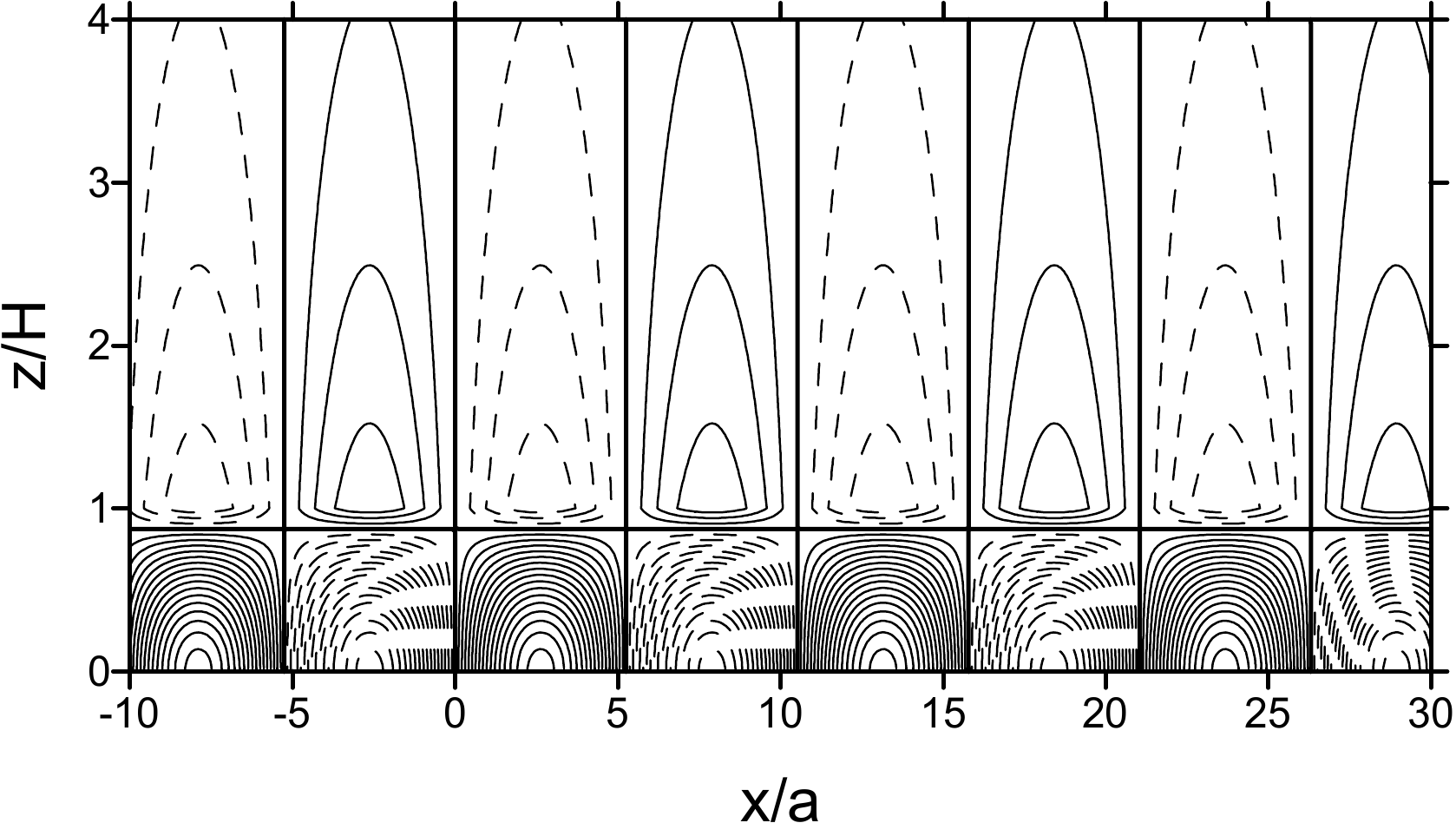}
  \caption{(a),(c),(e) Normalized vertical velocity perturbation $w/(U h_0/a)$ and (b),(d),(f) normalized horizontal velocity perturbation $u/(N_1 h_0)$ for the two-layer atmosphere of Scorer, for $l_2/l_1=0.2$, $l_1 H/\pi=0.6$ and $l_1 a=2$, from: (a),(b) FLEX numerical simulations; (c),(d) linear model with friction for $\lambda a/U=5 \times 10^{-4}$; (e),(f) inviscid linear theory. Note
that the results from linear theory in (e),(f) should be disregarded for $x<0$. Contour spacing: 0.2; Solid contours: positive values, dashed contours: negative values.}
\label{fig2}
\end{center}
\end{figure}
Before this is done, it is useful to check whether the theory, linear model with friction and numerical simulations are comparable. Figure \ref{fig2} shows the normalized vertical and horizontal flow perturbations associated with the waves generated for flow of the two-layer atmosphere of Scorer over a very small amplitude bell-shaped mountain, described by Equation (\ref{bellshape}). The dimensionless input parameters on which the normalized results of the linear theory depend are $l_2/l_1$, $l_1 H/\pi$ and $l_1 a$ \citep{Teixeira_etal_2013a}, which are expected to be the same as for the inviscid simulations of the FLEX model (since $l_1 h_0= 0.02$ is very small). It is assumed that $l_2/l_1=0.2$, $l_1 a=2$ and $l_1 H/\pi=0.6$. An additional input parameter in the linear model with friction is $\lambda U/a$. As can be seen, the behaviour of the 3 models for $x>0$ is very similar, with trapped lee waves totally dominating the flow (cf. Figure 18b of \citet{Teixeira_etal_2013a} for $l_1 H/\pi=0.8$ instead). The wavelength of the waves, and even the intensity of their velocity perturbations (evaluated by the number of contours) is quite similar between all cases, with the difference that the wave from inviscid linear theory is perfectly monochromatic, and so must be disregarded for $x<0$. The linear model with friction correctly suppresses the wave upstream of the mountain, albeit showing some differences in structure relative to the FLEX numerical simulation.

The most important message conveyed by Figure \ref{fig2} is, however, that the structure of the trapped lee wave is, for $x$ somewhat larger than 0 (say, $x/a>5$), almost indistinguishable between the 3 models. This corroborates the assumption underlying the calculations presented above that, for vanishing friction, it is appropriate to redefine the lower limit of integration in Equation (\ref{final}) as $0$. This is because an overwhelming contribution to the integral comes from substantially larger $x$, where the approximation from inviscid linear theory is very accurate, so the exact value of this lower integration limit, and the behaviour of the wave solution in its vicinity, are irrelevant. This result ultimately relies on the asymptotic approximation of \citet{Scorer_1949}, but its relevance for the specific purpose of evaluating the integral in Equation (\ref{final}) should be emphasized here.

\subsection{Inviscid results}

The extension of the results of \citet{Broad_2002} presented in Section \ref{method} will now be tested against numerical simulations. For this purpose, the inviscid linear model is compared with inviscid FLEX runs. Since, according to section \ref{method}, the horizontal flux of vertical momentum, defined with a $+\infty$ upper limit of integration, does not converge for purely inviscid flow, the definition using $x$ instead as upper limit of integration, included in Equation (\ref{constraint2}), is adopted here.

\begin{figure}[h!]
\begin{center}
 (a)\includegraphics[width=0.5\columnwidth]{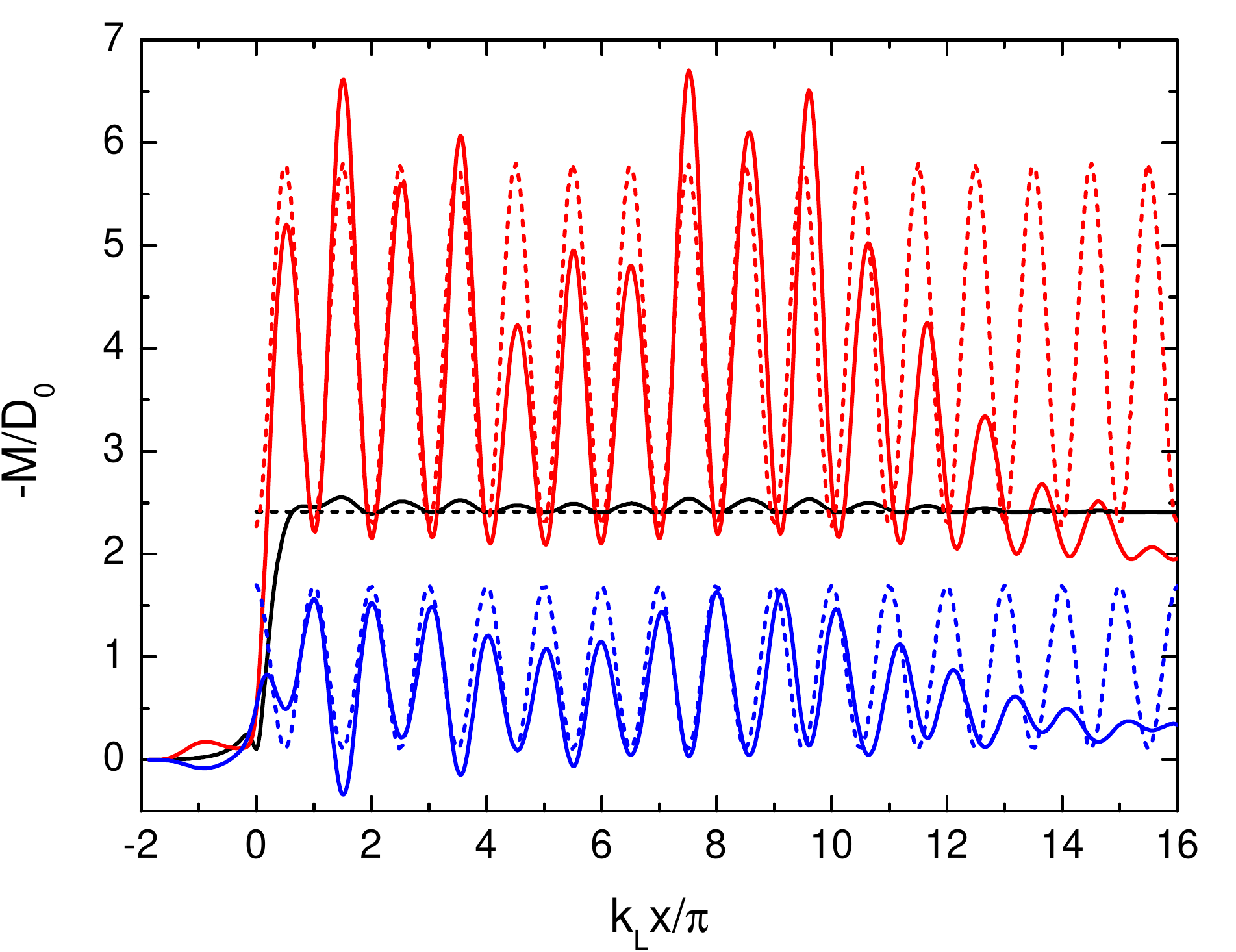}\\
 (b)\includegraphics[width=0.45\columnwidth]{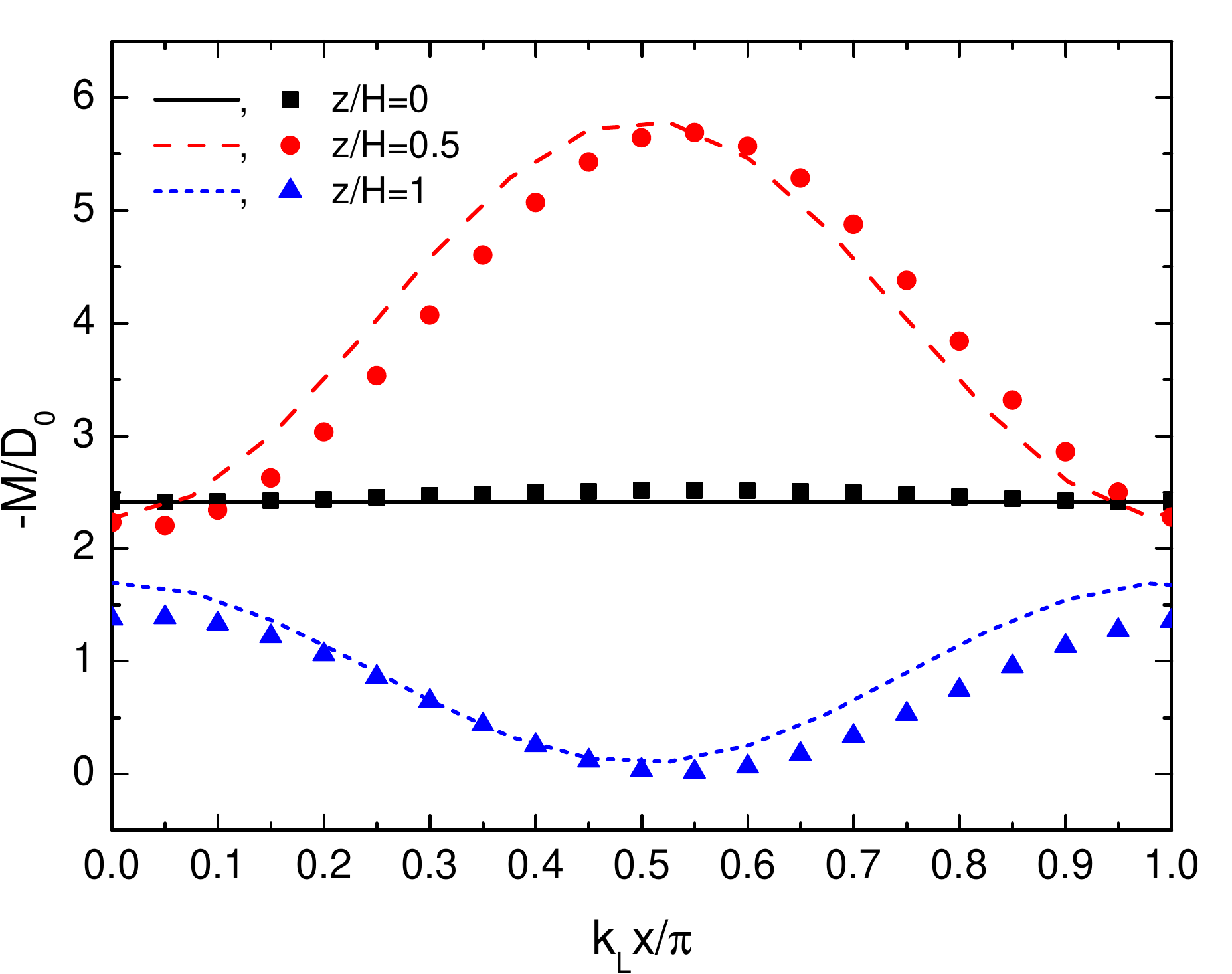}
 (c)\includegraphics[width=0.3\columnwidth]{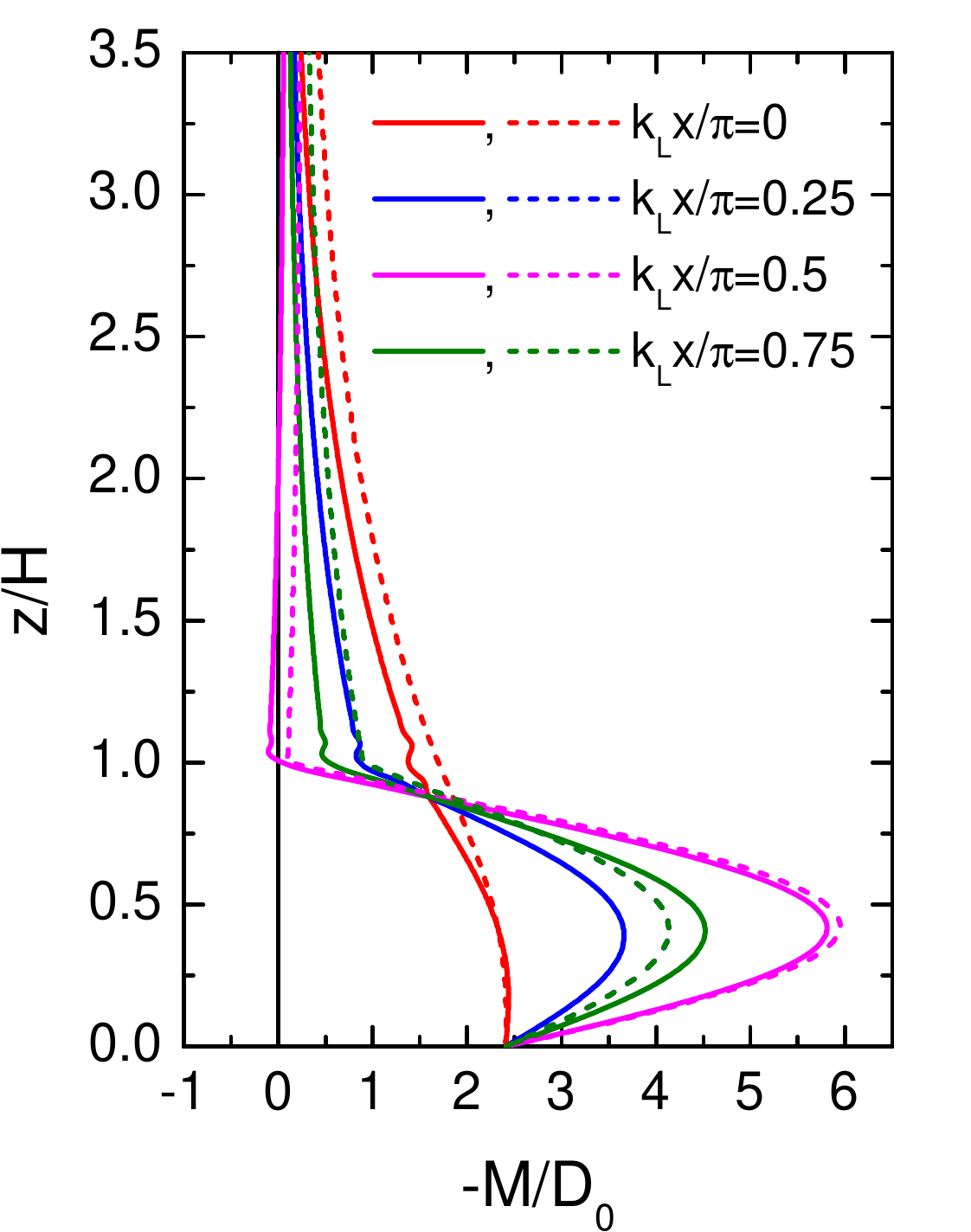}
  \caption{(a) Normalized momentum flux at 3 different levels from the FLEX model (solid lines) and from inviscid linear theory (Equations (\ref{broadmom})-(\ref{nonbroad})) (dashed lines), for $l_2/l_1=0.2$, $l_1 H/\pi=0.6$ and $l_1 a=2$. The horizontal coordinate is normalized using the theoretical wavenumber of the trapped lee waves, $k_L$. Black lines: $z/H=0$, red lines: $z/H=0.5$, blue lines: $z/H=1$. (b) Normalized momentum flux as a function of the wave phase, averaged over $1<k_L x/\pi<10$ from panel (a). Symbols: FLEX model, lines: inviscid linear model (see legend for details). (c) Profiles of the normalized momentum flux for key values of the wave phase (averaged as in panel (b)). Solid lines: FLEX model, dashed lines: inviscid linear theory. See legend for details (note that the dashed blue line coincides with the dashed green line and is hidden by it).
}
\label{fig3}
\end{center}
\end{figure}
Figure \ref{fig3} shows this momentum flux (corresponding to $M_B+M_{NB}$ defined according to Equations (\ref{broadmom})-(\ref{nonbroad}), but here also denoted by $M$, for convenience), normalized by the surface drag produced by hydrostatic waves in an atmosphere similar to the lower layer, but extending indefinitely, $D_0 = (\pi/4) \rho_0 U^2 l_1 h_0^2$ \citep{Teixeira_etal_2013a}, in 3 different ways. Figure \ref{fig3}a shows the momentum flux as a function of downstream distance at 3 different heights: $z/H=0, 0.5, 1$. $M/D_0$ oscillates with downstream distance (with an especially high amplitude at $z/H=0.5$) except at $z=0$. This latter result highlights the well-posedness of the inviscid surface drag problem, which \citet{Teixeira_etal_2013a} took advantage of, and is due to the fact that $z=0$ is the only height at which the trapped lee waves do not extend indefinitely, because of the surface boundary condition. Since the product of $u$ and $w$ is rather sensitive to phase differences in the oscillations, in addition to fluctuations in magnitude, the field of $-M/D_0$ from FLEX (solid lines) is not as regular in Figure \ref{fig3}a as those of $w$ and $u$ in Figure \ref{fig2}, showing some modulation, whose causes are unclear. Part of this modulation, particularly the monotonic amplitude decay existing towards the right edge of the domain, is due to the effect of the lateral sponge at the downstream boundary, but this effect appears to extend considerably beyond the space occupied by the sponge itself. Naturally, any amplitude modulation is totally absent in the results from inviscid linear theory (dashed lines). However, the overall magnitude of $-M/D_0$, amplitude of its oscillations, their wavelength and phase are quite in good agreement with the numerical simulations, particularly for $k_L x/\pi < 11$. This suggests averaging these fields over a number of wave cycles to make the comparison easier.

Figure \ref{fig3}b shows such an average, for the same heights as considered in Figure \ref{fig3}a, taken over the wave cycles existing between $k_L x/\pi=1$ and 10. The lower limit of $k_L x/\pi$ is dictated by the fact that the wave is not yet quasi-periodic for $0<k_L x/\pi<1$ and the upper limit by the decay of the wave towards the downstream boundary of the domain. It can be seen that the magnitude of $-M/D_0$, as well as the amplitude and phase of its oscillation over a wave cycle is in good agreement between FLEX and inviscid linear theory. The amplitude of the oscillation is a bit smaller in FLEX than in linear theory at $z/H=1$, and the oscillation is slightly out of phase at both $z/H=0.5,1$, owing to the slightly larger wavelength of the trapped lee waves in the numerical simulations (see Figure \ref{fig3}a).

In Figure \ref{fig3}c, full profiles of $-M/D_0$ are presented for key values of the wave phase. $k_L x/\pi=0$ is the point at which both the $u$ and $\zeta$ flow perturbations are zero and $w$ is a maximum (i.e. the reference point considered by \citet{Broad_2002}). At $k_L x/\pi=0.5$, $-M/D_0$ is in phase opposition to the preceding case, with the $u$ and $\zeta$ velocity perturbations being at their maxima and $w$ being zero. The intermediate phase points correspond to $k_L x/\pi=0.25,\,0.75$. Clearly, the most interesting results are for $k_L x/\pi=0,\,0.5$, as $-M/D_0$ behaves in an intermediate way for $k_L x/\pi=0.25,\, 0.75$.
When $k_L x/\pi=0$ (corresponding to the original calculations of \citet{Broad_2002}), $dM/dz=0$ at $z=0$ (a feature, that, as will be seen, is preserved in the quasi-inviscid results with vanishing friction), and $dM/dz$ is continuous at $z=H$. These `desirable' features may have influenced \citet{Broad_2002} to privilege this particular result. On the other hand, for $k_L x/\pi=0.5$, $-dM/dz$ is positive at $z=0$, with $-M/D_0$ attaining a maximum slightly below $z/H=0.5$ that is more than twice its value at $z=0$, and $dM/dz$ is discontinuous at $z=H$, with very small values of $-M/D_0$ in the upper layer. Agreement between the FLEX model and linear theory is good, especially considering the very large modulation that $-M/D_0$ undergoes over the wave cycle, but $-M/D_0$ from FLEX is slightly lower than from linear theory in the upper layer. There is a difference in the profiles of $-M/D_0$ at $k_L x/\pi=0.25,\,0.75$ from FLEX, which does not exist in those from linear theory, because of the phase difference between the two oscillations, shown in Figure \ref{fig3}b and commented on above.

Overall, it can be concluded that the FLEX model, run in inviscid mode, reproduces the main physical aspects of the structure of the momentum flux predicted by the inviscid linear theory presented before, which extends the analysis of \citet{Broad_2002}. The results emphasize that the vertical flux of horizontal momentum does not, in this case, take a well-defined form.

\subsection{Results with vanishing friction}

\begin{figure}[h!]
\begin{center}
  (a)\includegraphics[width=0.3\columnwidth]{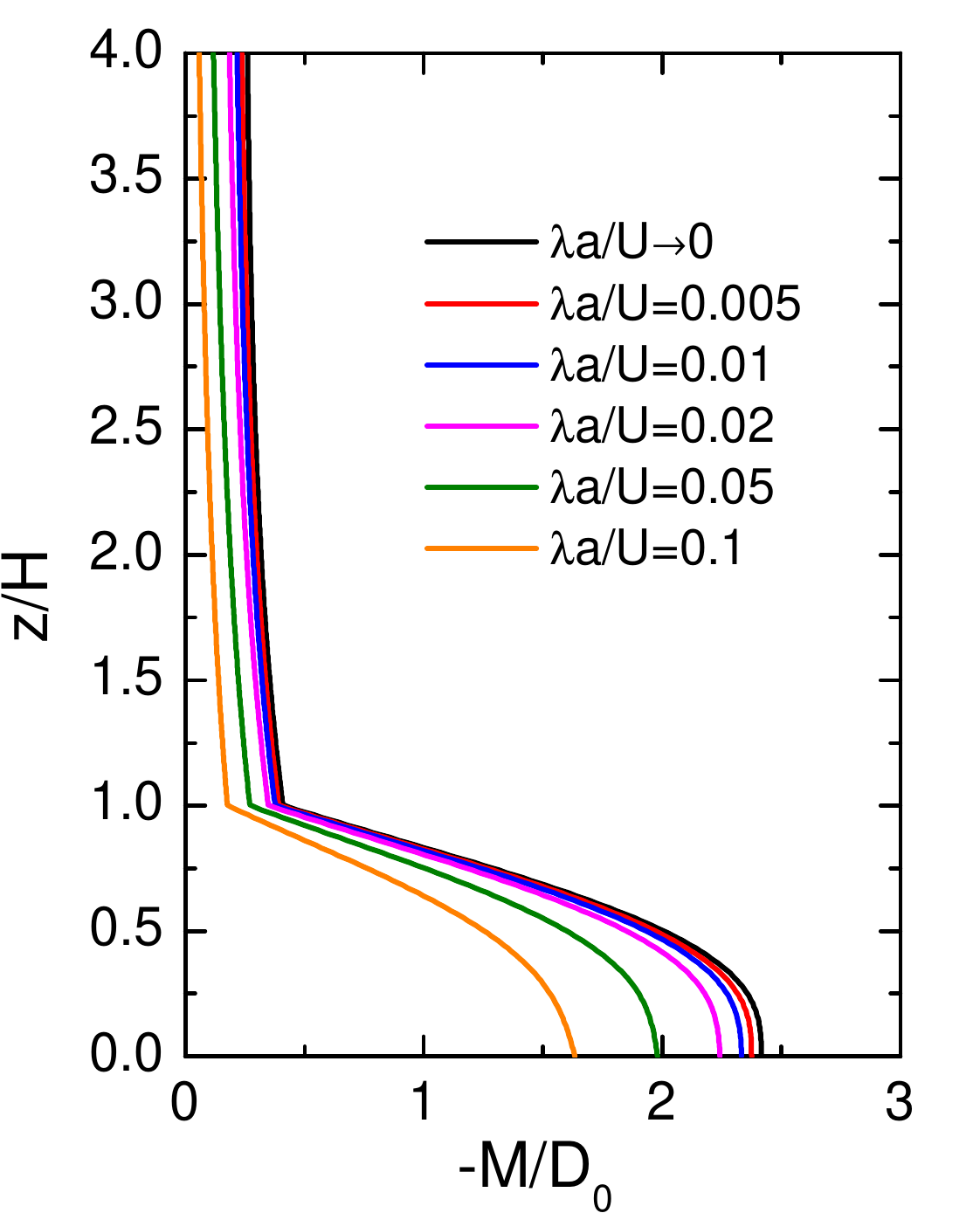}
  (b)\includegraphics[width=0.3\columnwidth]{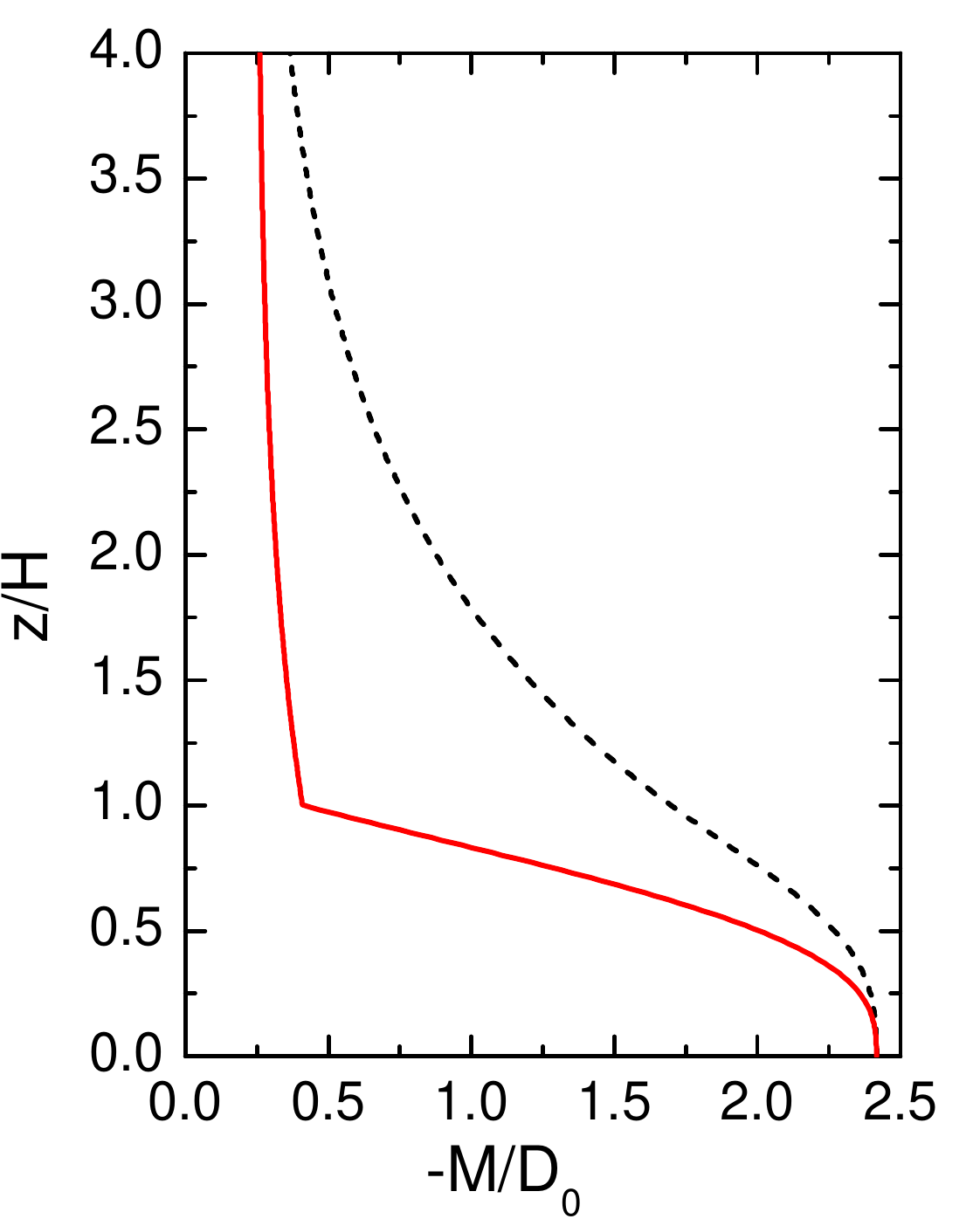}
  (c)\includegraphics[width=0.3\columnwidth]{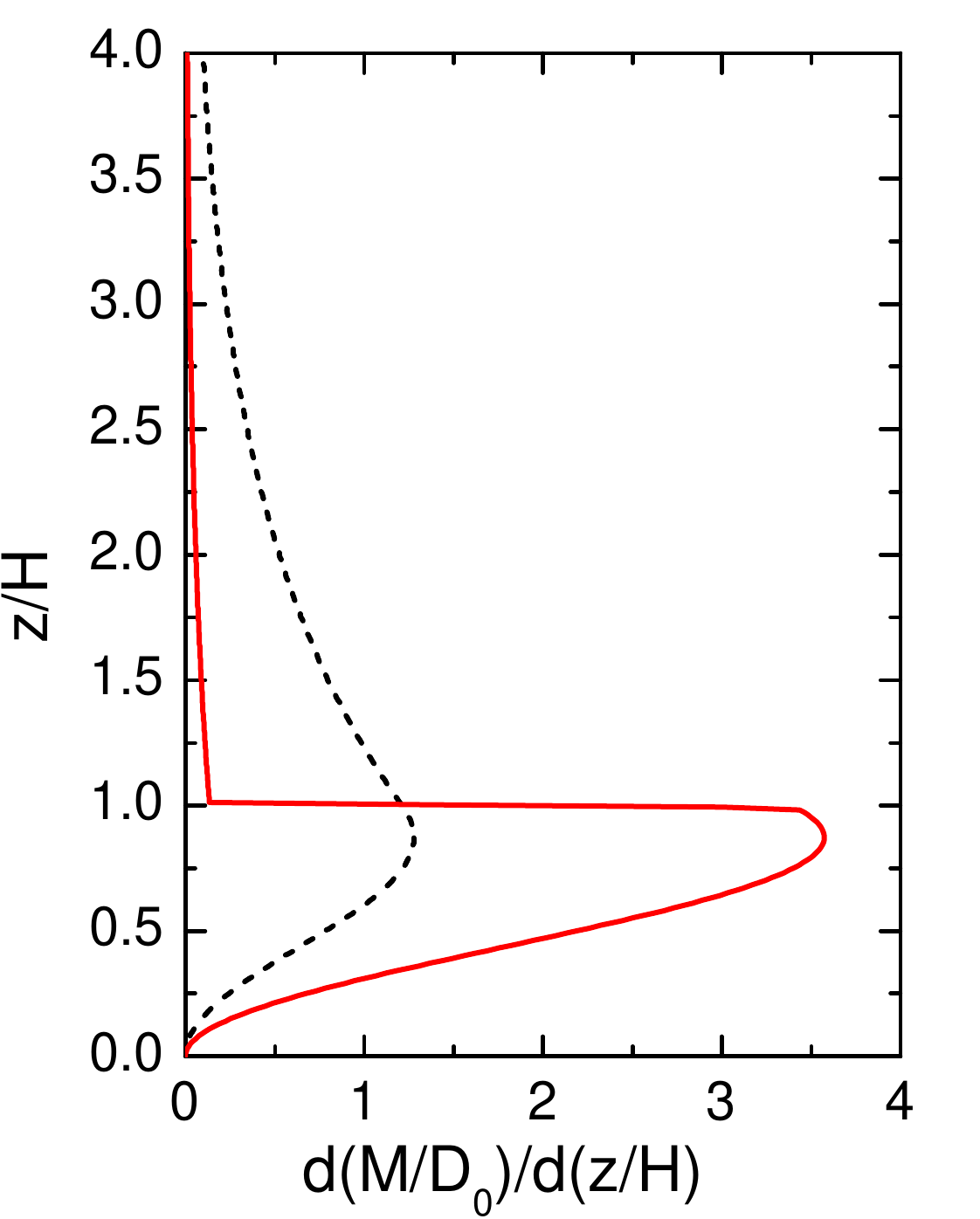}
\caption{(a) Normalized momentum flux profiles from quasi-inviscid linear theory (black line, corresponding to Equation (\ref{final4})) and from the linear model with friction (according to Equation (\ref{momfric})), for the two-layer atmosphere of Scorer with $l_2/l_1=0.2$, $l_1 H/\pi=0.6$ and $l_1 a=2$, for different values of the friction coefficient $\lambda a/U$ (see legend for details). (b) Momentum flux and (c) momentum flux divergence profiles, for the same input parameters, from Broad's theory, Equation (\ref{broadmom}) (black dashed line), and from quasi-inviscid theory, Equation (\ref{final4}) (red solid line).}
\label{fig4}
\end{center}
\end{figure}

Comparisons of inviscid simulations of the FLEX model with the quasi-inviscid linear theory (including vanishingly small friction) are presented next. First of all, it is necessary to ascertain that this linear theory does, indeed, represent the limit of very small, but non-zero friction, accurately. This is done in Figure \ref{fig4}a, which shows profiles of the normalized momentum flux from the quasi-inviscid linear theory (black line) and from the linear model with finite friction (colour lines), for $l_2/l_1=0.2$, $l_1 H/\pi=0.6$ and $l_1 a=2$ (the same conditions as considered in Figures \ref{fig2} and \ref{fig3}). In these results, the definition of $M$ includes the upper limit of integration $+\infty$, as is consistent with any non-zero friction (since the trapped lee waves necessarily decay downstream, no matter how slowly).

In Figure \ref{fig4}a, the momentum flux profiles are clearly very different from any of those presented in Figure \ref{fig3}c, including that from \citet{Broad_2002}. For sufficiently weak friction, $dM/dz=0$ at $z=0$, but $dM/dz$ is discontinuous at $z=H$, with $-M/D_0$ being much smaller in the upper layer than in the lower layer. Both aspects can be explained by differentiating Equation (\ref{final}) with respect to $z$, yielding
\begin{equation}
  \frac{\partial}{\partial z} \int_{-\infty}^{+\infty} u w \, dx = \frac{\lambda}{U} \left( N^2 - U \frac{d^2 U}{dz^2} \right) \int_{-\infty}^{+\infty} \zeta^2 dx,
  \label{finaldiff}
\end{equation}
If this equation is applied at $z=0$ it reduces to
\begin{equation}
 \left(\frac{\partial}{\partial z} \int_{-\infty}^{+\infty} u w \, dx \right)(z=0) = \frac{\lambda}{U(z=0)} \left\{ N^2(z=0) - U(z=0) \frac{d^2 U}{dz^2}(z=0) \right\} \int_{-\infty}^{+\infty} h^2 dx,
 \label{finaldiff1}
\end{equation}
where $\zeta(z=0)=h$ has been used. Since $h(x)$ does not depend on $\lambda$, the right-hand side of Equation (\ref{finaldiff1}) approaches 0 as $\lambda \rightarrow 0$ (note that this does not occur for $z>0$, because $\zeta^2$ then extends downstream for a distance proportional to $U/\lambda$). When $z>0$, Equation (\ref{finaldiff}) shows that $dM/dz$ is proportional to $N^2$ (if $d^2 U/dz^2=0$), and that is why in Figure \ref{fig4}a there is a discontinuity in $dM/dz$ at $z=H$, where $N$ is discontinuous. This is, however, a specific feature of Scorer's two-layer atmosphere, and would not exist for a more realistic static stability profile.

Figure \ref{fig4}a shows that the model with friction seamlessly approaches the quasi-inviscid theory as $\lambda a/U$ decreases to zero. In particular, it can be seen that in order for the $-M/D_0$ profile to be distinguishable between the two models, it is necessary that $\lambda a/U$ is substantially larger than the value assumed in Figure \ref{fig2}c,d. This corroborates that the quasi-inviscid linear theory is a consistent limit of the problem with friction as $\lambda \rightarrow 0$, with the advantage that it provides closed analytical expressions for the momentum flux (for this simplified atmosphere). A final aspect to note is that the non-zero value to which $-M/D_0$ asymptotes in the upper layer is due to the propagation of non-trapped waves, which are associated with a constant momentum flux profile (in the absence of any critical levels, as is the case). This non-zero constant is, however, by deliberate choice of the input parameters, a relatively small fraction of the total momentum flux, as this allows focusing primarily on the trapped lee waves. 

Figure \ref{fig4}b,c exemplifies how the results from quasi-inviscid theory, Equation (\ref{final4}), differ from those predicted by Broad's theory, Equation (\ref{broadmom}), for the same conditions as considered in Figure \ref{fig4}a. Although in Figure \ref{fig4}b the two results for $-M/D_0$ coincide at the surface and at high levels, they differ most in the region centred around $Z/H=1$, where Broad's model overestimates quasi-inviscid theory considerably. More importantly for drag parametrization, the divergence of the momentum flux shown in Figure \ref{fig4}c predicted by quasi-inviscid theory exceeds by a factor larger than 3 that predicted by Broad's theory immediately below the interface separating the two-layers, and is smaller by an even larger factor in the upper layer. Admittedly, this is a rather extreme example of this discrepancy, and the discontinuity in $d(M/D_0)/d(z/H)$ displayed by quasi-inviscid theory is, as pointed out above, an artifact of Scorer's two-layer atmosphere.

\begin{figure}[h!]
\begin{center}
  (a)\includegraphics[width=0.28\columnwidth]{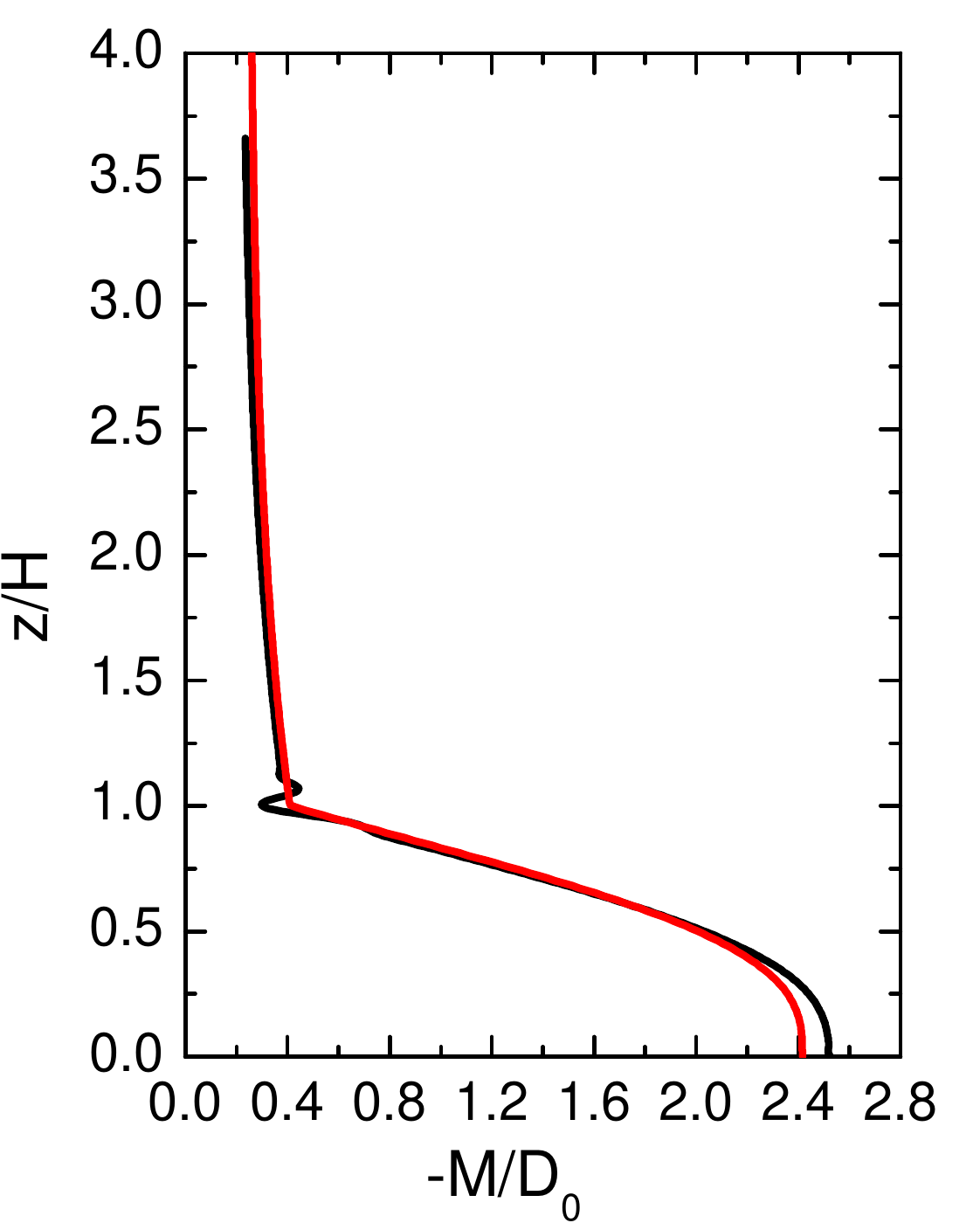}
  (b)\includegraphics[width=0.28\columnwidth]{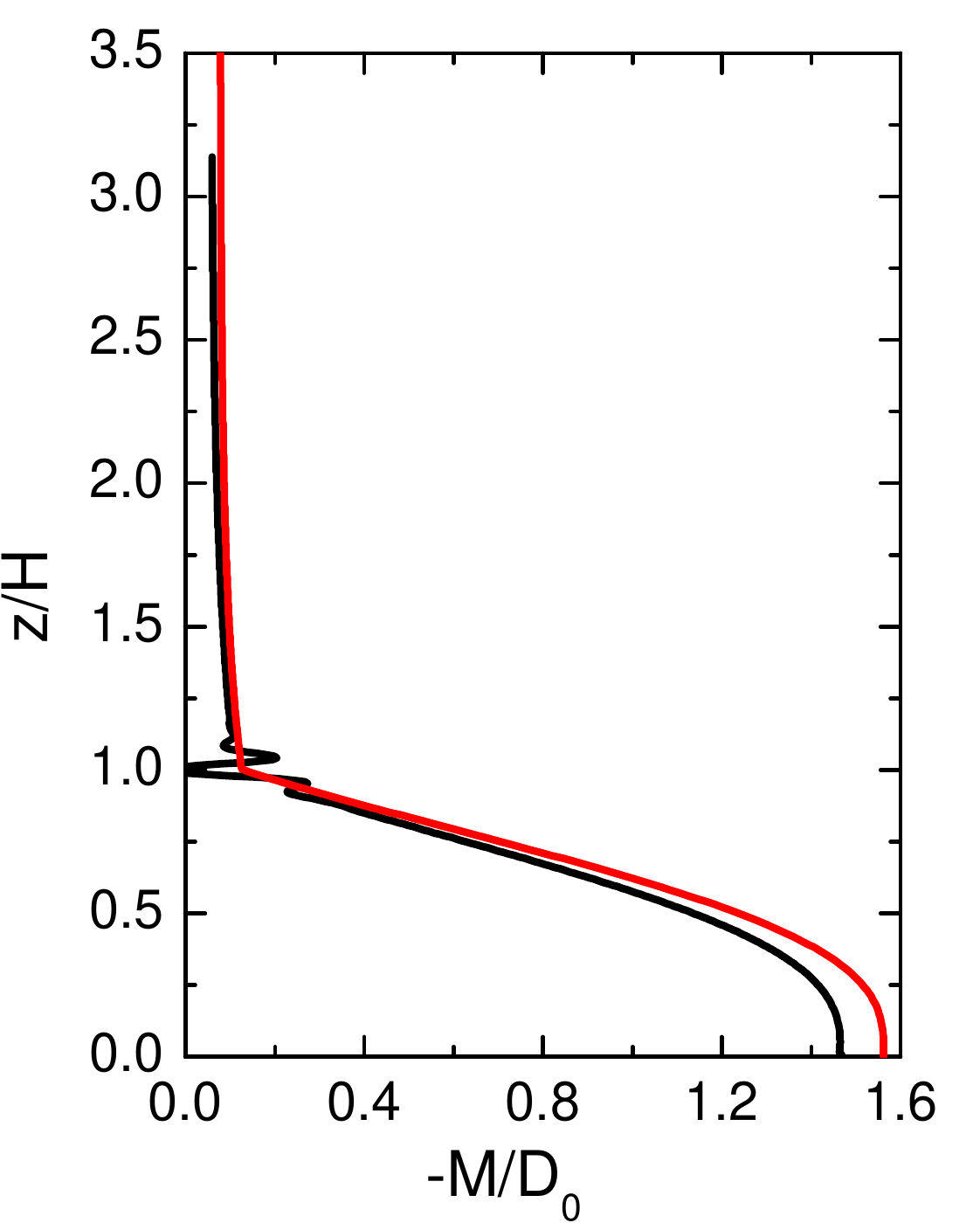}
  (c)\includegraphics[width=0.28\columnwidth]{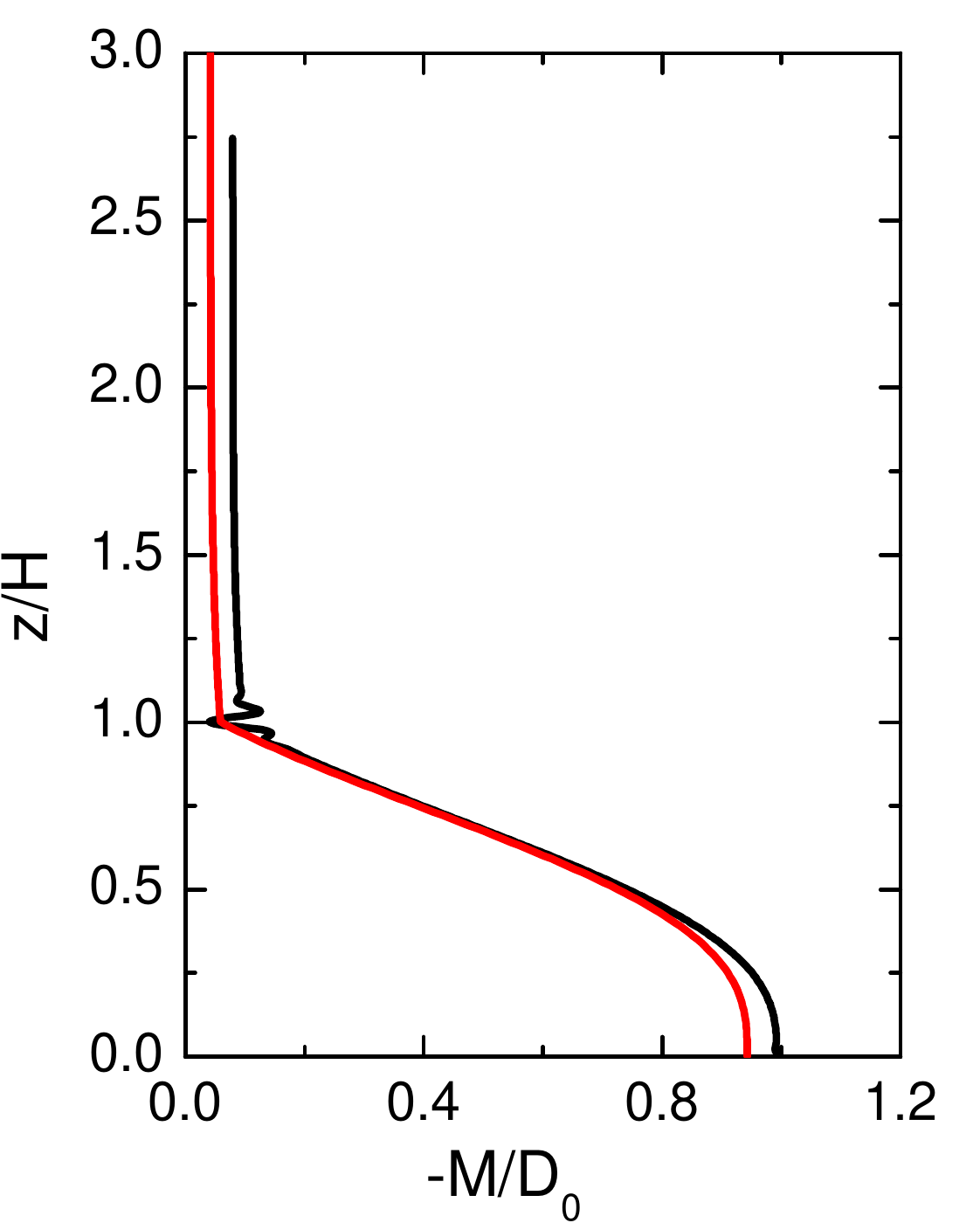}
  (d)\includegraphics[width=0.28\columnwidth]{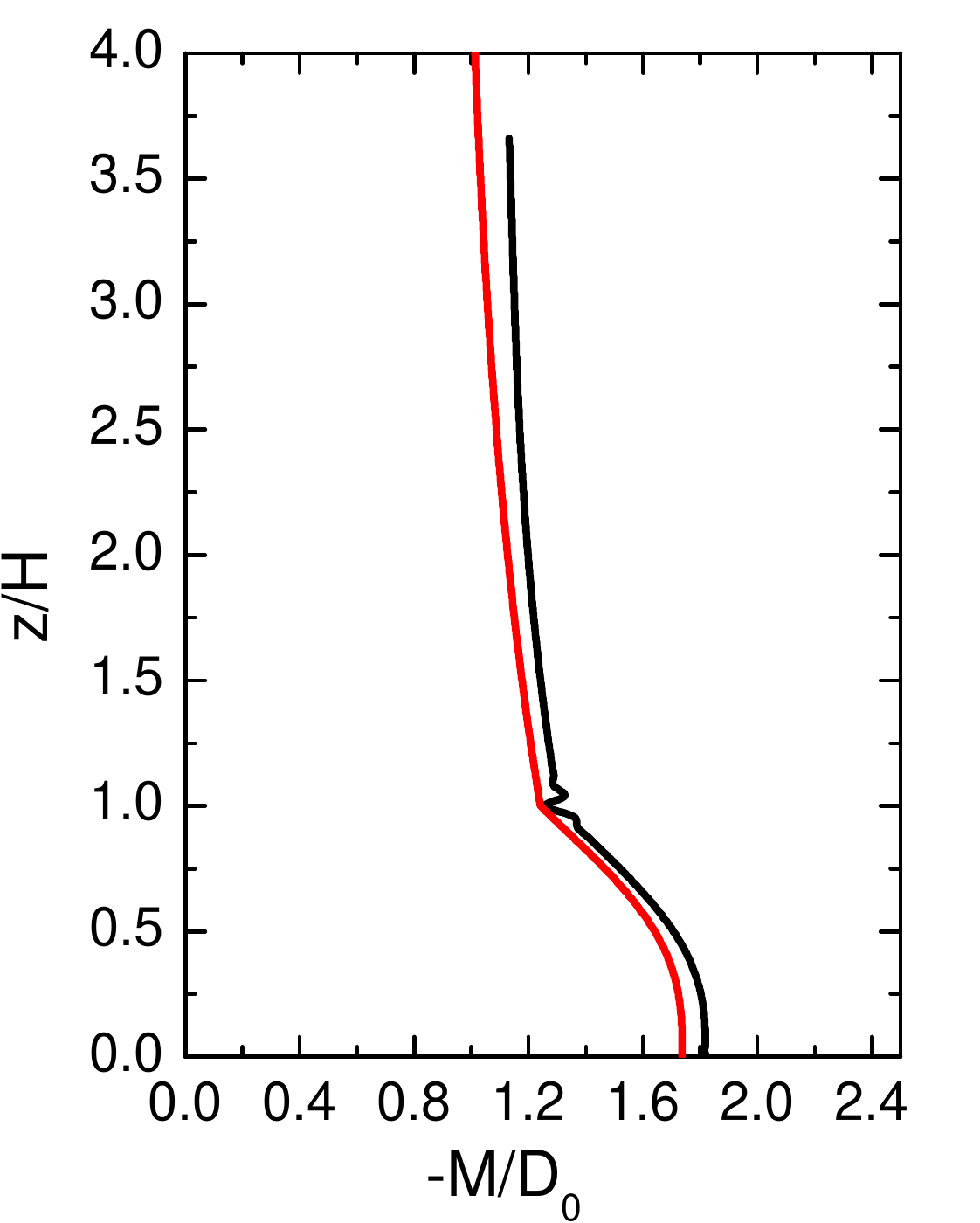}
  (e)\includegraphics[width=0.28\columnwidth]{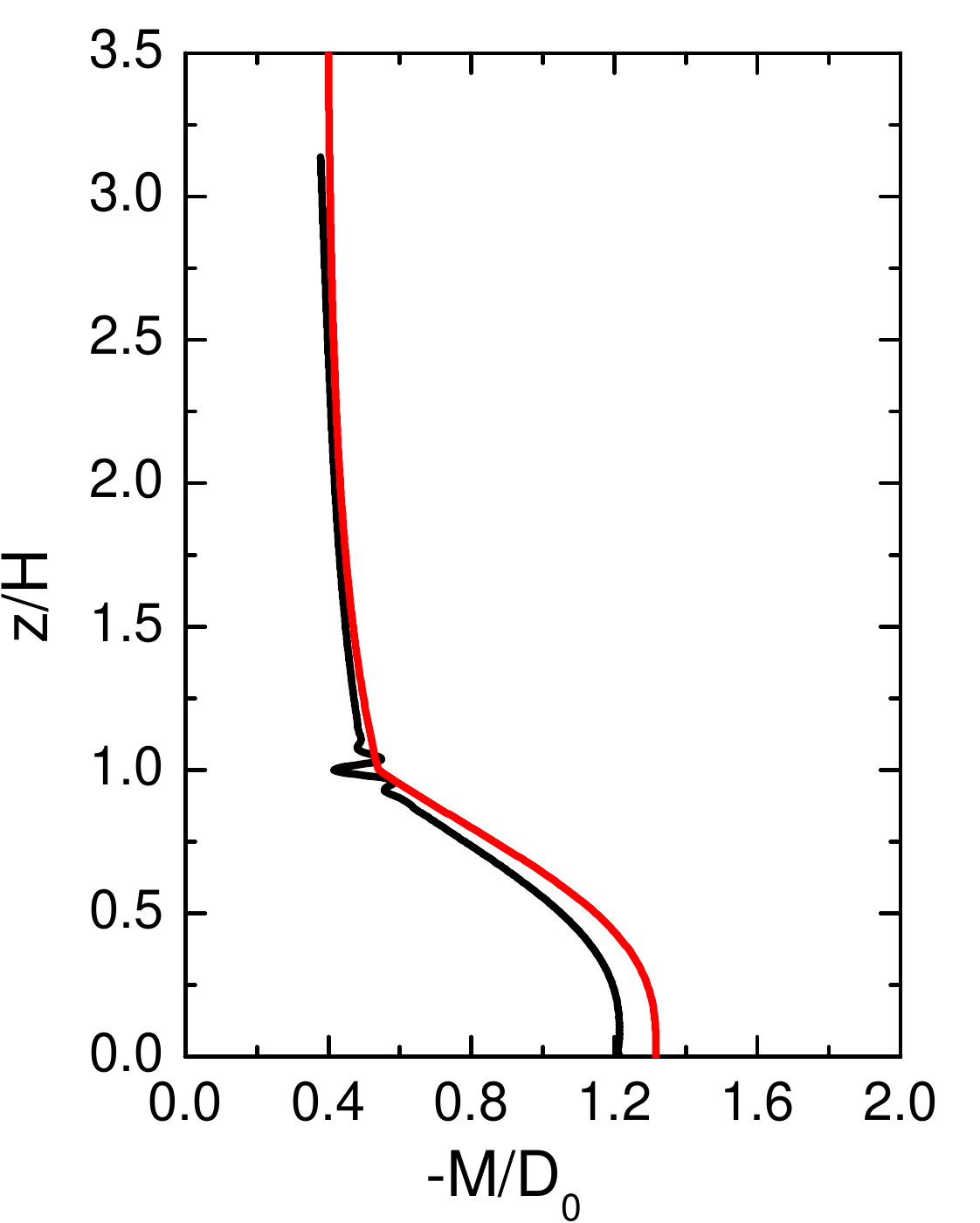}
  (f)\includegraphics[width=0.28\columnwidth]{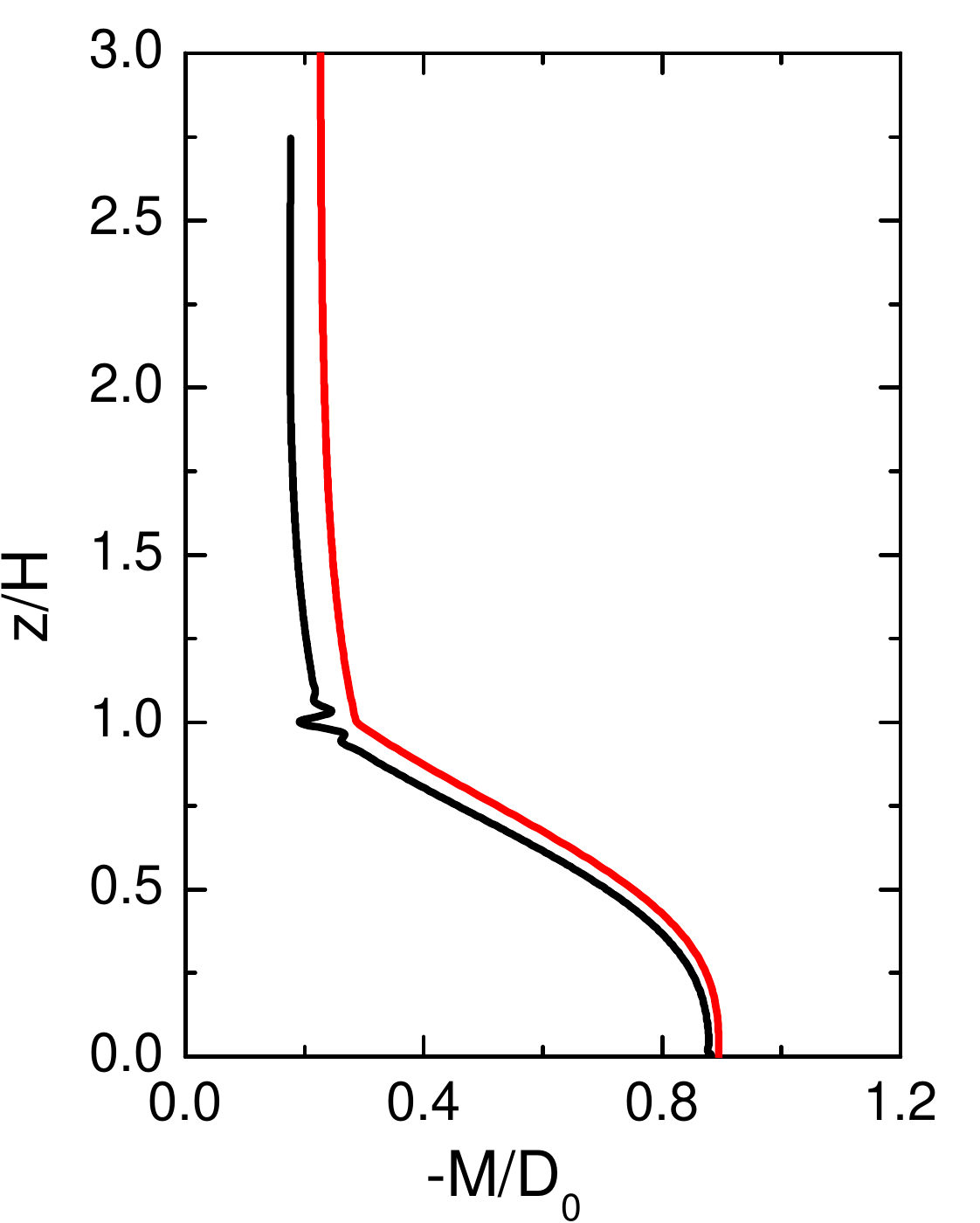}
  (g)\includegraphics[width=0.28\columnwidth]{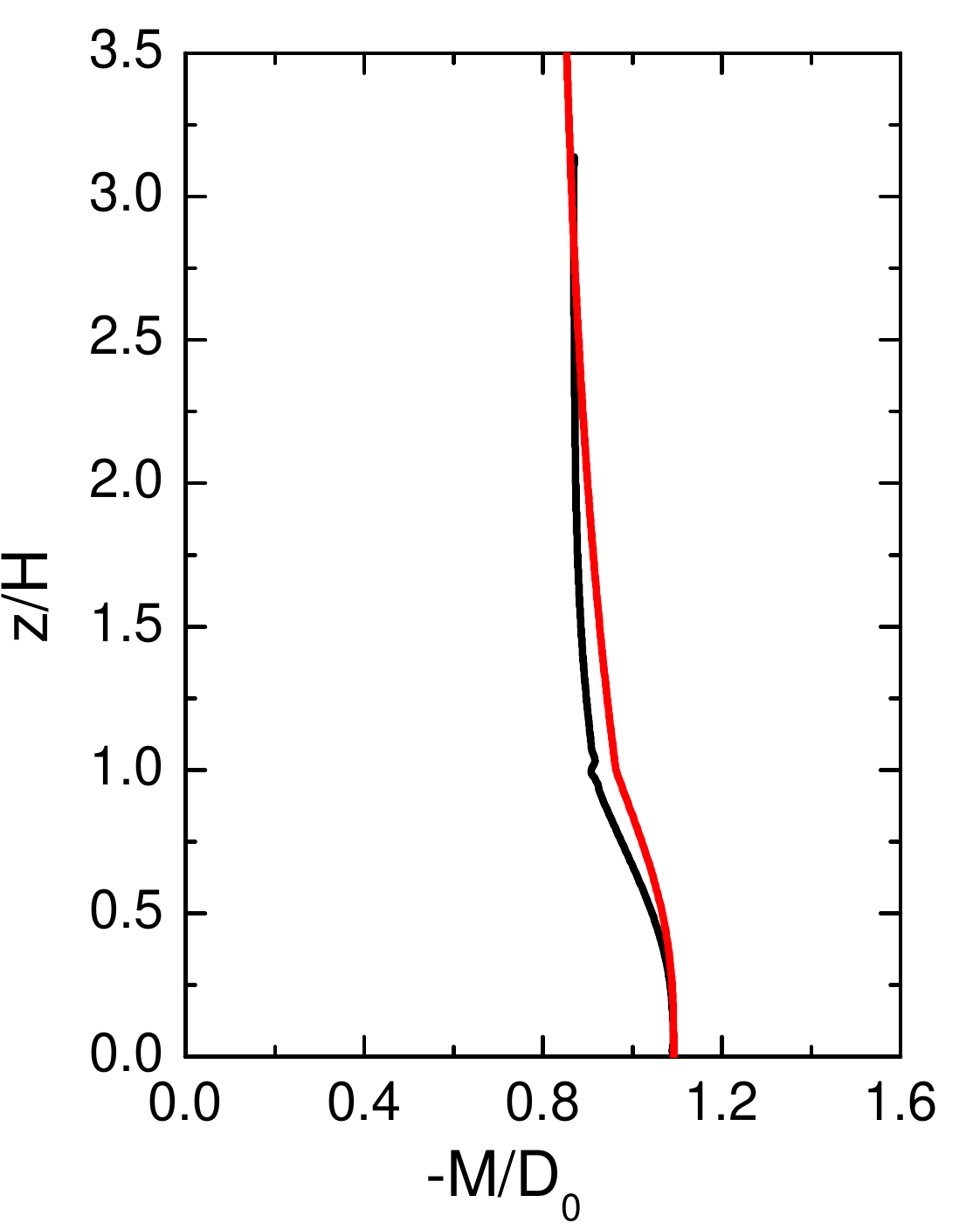}
  (h)\includegraphics[width=0.28\columnwidth]{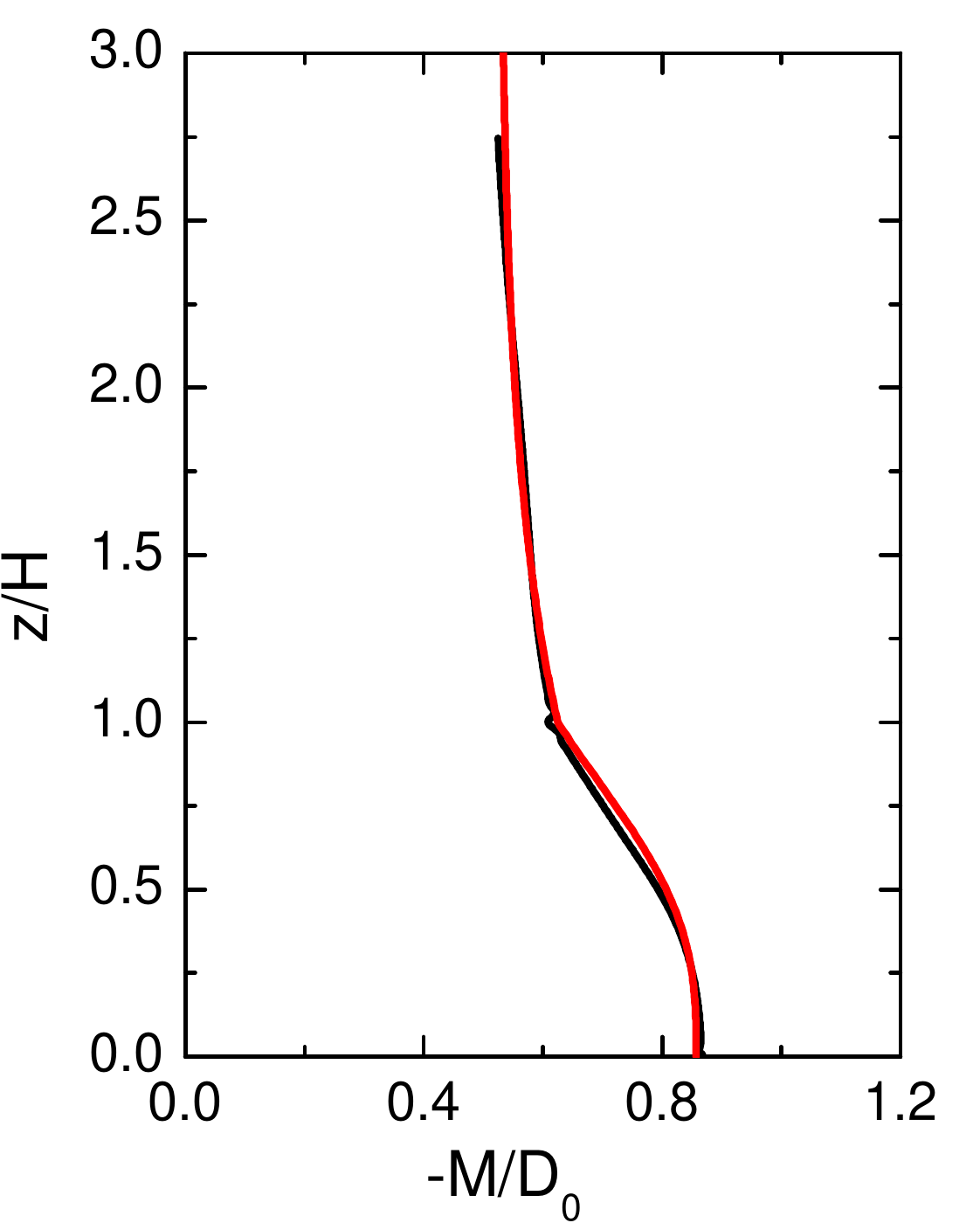}
  (i)\includegraphics[width=0.28\columnwidth]{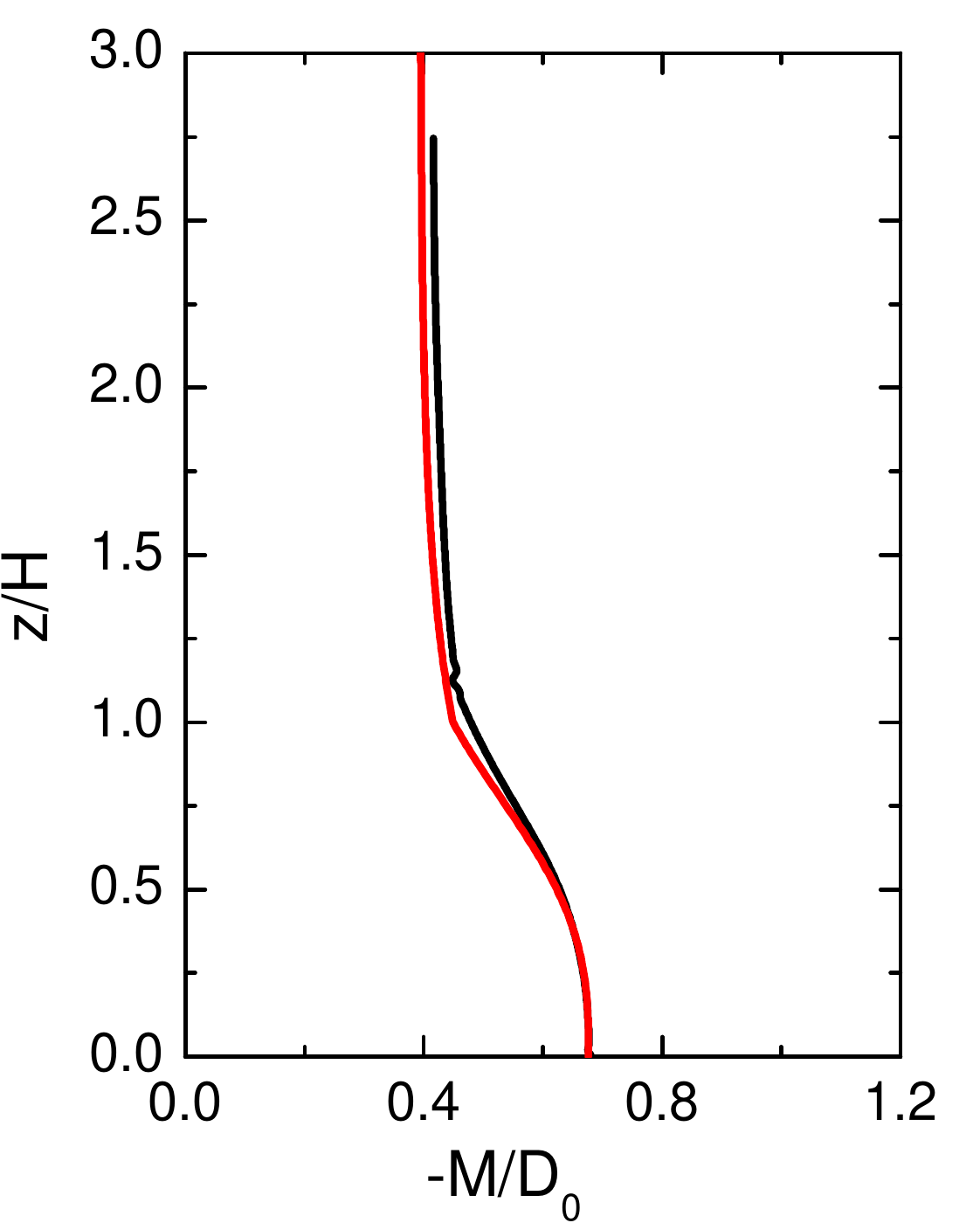}
\caption{Normalized momentum flux profiles for the two-layer atmosphere of Scorer for $l_1 a=2$ and: (a) $l_2/l_1=0.2$, $l_1 H/\pi=0.6$; (b) $l_2/l_1=0.2$, $l_1 H/\pi=0.7$; (c) $l_2/l_1=0.2$, $l_1 H/\pi=0.8$; (d) $l_2/l_1=0.4$, $l_1 H/\pi=0.6$; (e) $l_2/l_1=0.4$, $l_1 H/\pi=0.7$; (f) $l_2/l_1=0.4$, $l_1 H/\pi=0.8$; (g) $l_2/l_1=0.6$, $l_1 H/\pi=0.7$; (h) $l_2/l_1=0.6$, $l_1 H/\pi=0.8$; (i) $l_2/l_1=0.6$, $l_1 H/\pi=0.9$. Black lines: FLEX model, red lines: quasi-inviscid linear theory, Equation (\ref{final4}).
}
\label{fig5}
\end{center}
\end{figure}
In Figure \ref{fig5}, a comparison is made of a limited subset of momentum flux profiles for the two-layer atmosphere of Scorer, between the FLEX model and the quasi-inviscid linear theory.
To reproduce the conditions envisaged in theory as closely as possible, in FLEX the momentum fluxes are integrated horizontally over the full length of the computational domain, including the sponge damping layers.
Despite considering a range of values of $l_2/l_1$ and $l_1 H/\pi$, the results presented in Figure \ref{fig5} (like all the preceding results) are focused exclusively on the first trapped lee wave mode among the possible modes supported by Scorer's atmosphere (as can be checked against Figure 1 of \citet{Teixeira_etal_2013a} from the values of $l_2/l_1$ and $l_1 H/\pi$ assumed here). This choice is made because the first trapped lee wave mode is the strongest one and that likely to be represented most accurately in numerical simulations (as the wave reflection occurs closer to the ground, leading to less dispersive weakening of the wave), and also because the absence of additional wave modes makes the problem as `clean' as possible to illustrate the quasi-inviscid theory developed here. Of course, the theory is applicable to a much wider range of conditions. For $l_2/l_1=0.6$, the lowest value of $l_1 H/\pi$ is 0.7 instead of 0.6, because $l_1 H/\pi=0.6$ is theoretically expected to have zero trapped lee waves modes (by a narrow margin), and the focus here is on trapped lee waves. Values of $l_1 H/\pi$ adopted in Figure \ref{fig5} are also concentrated within the lower half of the interval spanned by $l_1 H/\pi$ corresponding to a single wave mode, because, as shown by \citet{Teixeira_etal_2013a}, the wavelength (Figure 6b of \citet{Teixeira_etal_2013a}) and the trapped lee wave amplitude (inferred from the corresponding drag -- Figure 6d,e of \citet{Teixeira_etal_2013a}), as well as the relative magnitude of the trapped lee wave drag compared with the drag of waves that propagate vertically into the upper layer, are all maximized for these conditions.

Figure \ref{fig5} confirms the results of quasi-inviscid theory with good accuracy, with a few exceptions in detail. Overall, both the surface value and shape of the profiles of $-M/D_0$ from FLEX and from the quasi-inviscid theory are in good agreement. Even in the cases where agreement is not perfect (surface value of $-M/D_0$ in Figure \ref{fig5}a-e, values of $-M/D_0$ in the upper layer in Figure \ref{fig5}c,d,f), the fractional difference is typically small, and qualitative agreement is very good.
In particular, the results from FLEX confirm the zero value of $dM/dz$ at the surface and its discontinuity at $z=H$ (where the momentum flux from FLEX displays some oscillations, presumably of numerical origin). This discontinuity obviously becomes weaker as $l_2/l_1$ increases, because $dM/dz$ was seen in Equation (\ref{final}) to be proportional to $N^2$. It is noteworthy that the magnitude of $-M/D_0$ at $z=0$ (which corresponds to the total surface drag) decreases both as $l_2/l_1$ and $l_1 H/\pi$ increase. The first result is due to the fact that the intensity of the wave reflection that generates the resonant trapped lee wave is proportional to the contrast in static stability between the two layers. The second result is due to the fact that, for a given mode, the intensity of the trapped lee wave (of which the associated drag is a good measure) decreases as $l_1 H/\pi$ increases (this can be seen in Figure 6d of \citet{Teixeira_etal_2013a}). As noted above, this is due to dispersion effects. The magnitude of the drag associated with waves that propagate vertically in the upper layer (shown by the magnitude of $-M/D_0$ near the top of the domain displayed in Figure \ref{fig5}) does not behave as monotonically with the input parameters. It is clear that these waves become weaker (as diagnosed by their momentum flux) when $l_1 H/\pi$ increases, but their variation with $l_2/l_1$ is not as obvious. It can be noticed, however, that in relative terms, these waves become more important compared to the trapped lee waves (as shown by the relative importance of their momentum flux, which is constant with height), as $l_2/l_1$ increases, since they do not require reflection at a layer to exist, unlike trapped lee waves.

One aspect deserving a more detailed analysis is that, while in quasi-inviscid theory the decay of the trapped lee waves with distance downstream (which enabled the calculation of the momentum flux from Equation (\ref{final4})) is exponential (see Equation (\ref{defzeta})), in the FLEX numerical model it takes quite a different form. Although the wave field, and especially the partially integrated momentum flux, is rather noisy (as shown in Figure \ref{fig3}), in the part of the computational domain outside the lateral sponges the FLEX model is nominally inviscid, so the amplitude of the trapped lee wave should be roughly constant. When the downstream lateral sponge is reached, the wave field decays to zero in some way prescribed by the sponge damping. Since Equation (\ref{final4}) was derived assuming an exponential decay, the fact that the FLEX results are so close to those of the quasi-inviscid theory is puzzling. The explanation may be in the more general form of the momentum flux expressed by Equation (\ref{final2}). In that equation, clearly if the factor modulating the amplitude along $x$ (an exponential in this case, but it could be any other function that decays to zero as $x \rightarrow +\infty$) does not depend on $z$, the shape of the profile of $M$ is determined only by the integral in the $z$ direction. The integral in the $x$ direction just plays the role of limiting the magnitude of the term on the right-hand side of Equation (\ref{final2}), being therefore equivalent to a scaling factor. It can thus be argued that, for any type of downstream decay of the trapped lee wave caused by weak friction that does not depend on $z$, the result produced by Equation (\ref{final2})
still holds. This endows the present results with considerable generality, making them more relevant.
The results also suggest that dissipation in the sponge layer existing at the downstream boundary of the domain in the FLEX simulations can be considered weak (as this is one of the assumptions in quasi-inviscid theory). This is consistent with the requirement that wave decay in the sponge layer be sufficiently gradual to avoid upstream reflections. It also seems likely that the assumption of weak dissipation is satisfied often in Nature, as suggested by the considerable spatial extent of many observed trapped lee wave patterns.

\section{Conclusions}

This study presents a long overdue new theory for the momentum fluxes associated with trapped lee waves, whose divergence corresponds to the drag exerted by mountains on the atmosphere, mediated by the waves. As a first approximation, linear 2D trapped lee waves were considered, to build the necessary theoretical framework under the most basic assumptions. Friction, which needs to be taken into account, was included in the simplest possible form, as a Rayleigh damping applied only to the momentum equations. The calculations were developed in the limit of vanishing friction (a Rayleigh damping coefficient $\lambda$ approaching zero). The solutions to the trapped lee waves and associated wave momentum flux problem were found to be self-consistent in this limit, providing a simplified framework that allows maximally general analytical results to be obtained.

The wave momentum flux was found to be expressed in terms of the product of $\lambda$ and an integral in space (in the horizontal and vertical directions) involving a quadratic quantity of the wave field: the square of the vertical streamline displacement. This integral increases in inverse proportion to $\lambda$ (because the waves then extend for a larger distance before they decay). So, despite the fact that the momentum flux is written in terms of an expression involving $\lambda$, this dependence cancels out in the limit $\lambda \rightarrow 0$, yielding a quasi-inviscid approximation that is well-posed mathematically, finite, and independent of $\lambda$. Although the details of this result rely on the adopted Rayleigh damping formulation for friction, the underlying logic extends to other forms of friction. It is plausible, for example, that the same arguments would qualitatively apply to momentum fluxes calculated with the diffusive representation of friction adopted by \citet{Soufflet_etal_2022}.

The results were illustrated by application to the two-layer atmosphere of Scorer, where wave trapping is induced by a piecewise constant profile of static stability, with a discontinuity at the top of the trapping layer. For this atmosphere, the results mentioned above about the independence of the momentum flux profile from $\lambda$ in the limit $\lambda \rightarrow 0$ were explicitly confirmed. It was possible to derive a closed-form analytical expression for the momentum flux profile, which is different from any of the expressions for the partial momentum flux (i.e. only extending up to a certain part of the trapped lee wave train downstream of the mountain) that could be obtained by extension of the perfectly inviscid theory of \citet{Broad_2002}. As a general result and for quasi-inviscid conditions, the momentum flux divergence is zero at the ground, as originally predicted by \citet{Broad_2002}, although, of course, boundary layer effects are likely to modify the behaviour of the momentum flux near the surface \citep{Turner_etal_2021}. Specifically for Scorer's atmosphere, there is a discontinuity in the momentum flux divergence at the interface between the two layers, since that divergence is proportional to the static stability $N^2$. This yields momentum flux divergence profiles that are quite different from those predicted by Broad's theory (which are continuous at this interface), corresponding to a larger drag exerted on the atmosphere near the top of the lower layer, and much lower drag within the upper layer. For Scorer's atmosphere, the momentum flux associated with waves that propagate vertically into the upper layer has no divergence, but its magnitude varies according to the relative importance of those waves and trapped lee waves. In the present study, cases where trapped lee waves are dominant have been selected, since these are totally responsible for the momentum flux divergence at low levels. One aspect on which the present quasi-inviscid theory and that of \cite{Broad_2002} agree is the magnitude of the momentum flux at $z=0$ (which coincides with the total surface drag for weak friction). This is a consequence of the fact that the problem of trapped lee wave drag is well-posed mathematically, even for perfectly inviscid flow (\citealp{Teixeira_etal_2013a}; \citealp{Teixeira_etal_2013b}), since contributions to this drag are confined to the vicinity of the isolated orography generating the trapped lee waves, even if the waves themselves extend indefinitely downstream.

The present preliminary results
should be viewed as an important contribution to the establishment of a workable
theory of the momentum fluxes produced by trapped lee waves and their impact on the atmosphere. However, what the theory, as described in this study, provides, is only the vertical flux of horizontal momentum integrated over the total area (in this 2D case, distance) spanned by the waves. Since trapped lee waves can extend over quite a large area (or long distance) downstream of their source, to know their integrated effect as a  function of height (which is what is provided here) is highly relevant, but not the whole story. It would also be useful to obtain the local impact of the trapped lee waves on the mean flow at given points within the wave field, as that region is likely to occupy a substantial range of model grid-points in reasonably high-resolution weather prediction numerical simulations. A treatment of this aspect would require knowing not only the vertical flux of horizontal momentum at each point (or at least over a smaller finite area), but also the horizontal flux of horizontal momentum (since, in the middle of the trapped lee wave field, the latter is not zero). \citet{Xue_Giorgietta_2021} and \citet{Xue_etal_2022}, for example, attempted to interpret their numerical simulations of trapped lee waves applying momentum budgets locally, but, as pointed out above, they used for that purpose purely inviscid theory. An extension of the present quasi-inviscid theory would allow this to be done in a more physically consistent way.

The Rayleigh damping approach adopted to represent friction in the present study may be viewed as non-optimal, due to its crudeness and specificity. However, the independence of the quasi-inviscid results from $\lambda$ and their agreement with the inviscid numerical simulations suggest that they may be more general than expected. Except within the atmospheric boundary layer, friction in a stratified atmosphere is typically quite weak, and this is also corroborated by the large horizontal extent of trapped lee waves that can be visualized through cloud condensation in satellite images. For these reasons, there is scope to believe that the results presented here may constitute a
good basis for the development of new physically-based drag parametrizations for trapped lee waves.

\bibliography{references}

\end{document}